\documentclass[twocolumn,twocolappendix]{aastex631}
\usepackage{amsmath,textcomp,gensymb}

\shortauthors{Behmard et al.}
\graphicspath{{./}{figures/}}
\begin{document}

\title{A Data-Driven M Dwarf Model and Detailed Abundances for $\sim$17,000 M Dwarfs in SDSS-V}

\author[0000-0003-0012-9093]{Aida Behmard}
\affiliation{Center for Computational Astrophysics, Flatiron Institute, 162 Fifth Ave, New York, NY 10010, USA; abehmard@flatironinstitute.org}
\altaffiliation{Flatiron Research Fellow}
\affiliation{American Museum of Natural History, 200 Central Park West, Manhattan, NY 10024, USA}

\author[0000-0001-5082-6693]{Melissa K. Ness}
\affiliation{Research School of Astronomy and Astrophysics, Australian National University, Canberra, ACT 2601, Australia}

\author[0000-0003-0872-7098]{Andrew R. Casey}
\affiliation{School of Physics and Astronomy, Monash University, Melbourne, VIC 3800, Australia}
\affiliation{ARC Centre of Excellence for All Sky Astrophysics in Three Dimensions (ASTRO-3D), Australia}
\affiliation{Center for Computational Astrophysics, Flatiron Institute, 162 Fifth Ave, New York, NY 10010, USA}

\author[0000-0003-4540-5661]{Ruth Angus}
\affiliation{American Museum of Natural History, 200 Central Park West, Manhattan, NY 10024, USA}

\author[0000-0001-6476-0576]{Katia Cunha}
\affiliation{University of Arizona, Tucson, AZ 85719, USA}
\affiliation{Observatório Nacional, São Cristóvão, Rio de Janeiro, Brazil}

\author[0000-0002-7883-5425]{Diogo Souto}
\affiliation{Departamento de F\'isica, Universidade Federal de Sergipe, Av. Marcelo Deda Chagas, S/N Cep 49.107-230, S\~ao Crist\'ov\~ao, SE, Brazil}

\author[0000-0003-4769-3273]{Yuxi(Lucy) Lu}
\affiliation{Department of Astronomy, The Ohio State University, Columbus, 140 W 18th Ave, OH 43210, USA}
\affiliation{Center for Cosmology and Astroparticle Physics (CCAPP), The Ohio State University, 191 W. Woodruff Ave., Columbus, OH 43210, USA}

\author[0000-0001-7258-1834]{Jennifer A. Johnson}
\affiliation{Department of Astronomy and Center for Cosmology and AstroParticle Physics, Ohio State University, Columbus, OH 43210, USA}

%% Note that the \and command from previous versions of AASTeX is now
%% depreciated in this version as it is no longer necessary. AASTeX 
%% automatically takes care of all commas and "and"s between authors names.

%% AASTeX 6.31 has the new \collaboration and \nocollaboration commands to
%% provide the collaboration status of a group of authors. These commands 
%% can be used either before or after the list of corresponding authors. The
%% argument for \collaboration is the collaboration identifier. Authors are
%% encouraged to surround collaboration identifiers with ()s. The 
%% \nocollaboration command takes no argument and exists to indicate that
%% the nearby authors are not part of surrounding collaborations.

%% Mark off the abstract in the ``abstract'' environment. 
\begin{abstract}
The cool temperatures of M dwarf atmospheres enable complex molecular chemistry, making robust characterization of M dwarf compositions a long-standing challenge. Recent modifications to spectral synthesis pipelines have enabled more accurate modeling of M dwarf atmospheres, but these methods are too slow for characterizing more than a handful of stars at a time. Data-driven methods such as \emph{The Cannon} are viable alternatives, and can harness the information content of many M dwarfs from large spectroscopic surveys. Here, we train \emph{The Cannon} on M dwarfs with FGK binary companions from the Sloan Digital Sky Survey-V/Milky Way Mapper (SDSS-V/MWM), with spectra from the Apache Point Observatory Galactic Evolution Experiment (APOGEE). The FGK-M pairs are assumed to be chemically homogeneous and span $-$0.56 $<$ [Fe/H] $<$ 0.31 dex. The resulting model is capable of inferring M dwarf $T_{\textrm{eff}}$ and elemental abundances for Fe, Mg, Al, Si, C, N, O, Ca, Ti, Cr, and Ni with median uncertainties of 13 K and 0.018$-$0.029 dex, respectively. We test the model by verifying that it reproduces reported abundance values of M dwarfs in open clusters and benchmark M dwarf datasets, as well as expected metallicity trends from stellar evolution. We apply the model to 16,590 M dwarfs in SDSS-V/MWM and provide their detailed abundances in our accompanying catalog.

%We introduce a new implementation of \emph{The Cannon} that corrects for differential diffusion effects between FGK and M dwarf binary companions, and use M dwarfs with diffusion-corrected
\end{abstract}


%% Keywords should appear after the \end{abstract} command. 
%% The AAS Journals now uses Unified Astronomy Thesaurus concepts:
%% https://astrothesaurus.org
%% You will be asked to selected these concepts during the submission process
%% but this old "keyword" functionality is maintained in case authors want
%% to include these concepts in their preprints.
%\keywords{Classical Novae (251) --- Ultraviolet astronomy(1736) --- History of astronomy(1868) --- Interdisciplinary astronomy(804)}

%% From the front matter, we move on to the body of the paper.
%% Sections are demarcated by \section and \subsection, respectively.
%% Observe the use of the LaTeX \label
%% command after the \subsection to give a symbolic KEY to the
%% subsection for cross-referencing in a \ref command.
%% You can use LaTeX's \ref and \label commands to keep track of
%% cross-references to sections, equations, tables, and figures.
%% That way, if you change the order of any elements, LaTeX will
%% automatically renumber them.
%%
%% We recommend that authors also use the natbib \citep
%% and \citet commands to identify citations.  The citations are
%% tied to the reference list via symbolic KEYs. The KEY corresponds
%% to the KEY in the \bibitem in the reference list below. 

\section{Introduction} \label{sec:intro}
M dwarfs are the most common stars, comprising $\sim$70\% of the Galactic stellar population \citep{miller1979,bochanski2010,henry2018}. Their low masses translate to slow hydrogen fusion rates at their cores, resulting in long main-sequence lifetimes that exceed the age of the universe (e.g., \citealt{woolf2020}). Because M dwarfs are often old, their chemical compositions encode nucleosynthetic processes and interstellar medium enrichment from early generations of higher mass stars. M dwarfs are thus fossil records of Galactic chemical evolution \citep{bochanski2010,woolf2012}. Additionally, M dwarfs are ideal for exoplanet detection and characterization. Their small sizes and low masses lead to strong transit and radial velocity signals, even for small, cool planets that approach the Earth-like regime \citep{nutzman2008,trifonov2018}. This makes M dwarfs especially attractive targets for exoplanet surveys, e.g., the Transiting Exoplanet Survey Satellite (TESS) and PLAnetary Transits and Oscillations of stars (PLATO) missions \citep{plato2024}, and for planetary atmosphere investigations with JWST (e.g., \citealt{clampin2008,muirhead2018}). Planet properties gain valuable context with knowledge of host star chemistry, which reflect the compositions of planet building block material from protoplanetary disks. Thus, robust methods for measuring M dwarf chemistry will revolutionize our understanding of both planet formation and the assembly history of our galaxy.

Constraining M dwarf chemical compositions is notoriously difficult. Traditional spectroscopic methods that rely on physical stellar atmosphere models are optimized for solar-like stars with temperatures above $\sim$4500 K (e.g., \citealt{brewer2018,jonsson2020,hayes2022}). M dwarfs are cooler, so their atmospheres can harbor molecules (e.g.,  TiO, VO, MgH, CaH, FeH, H$_{2}$O, CO) that create dense clusters of molecular lines in the optical and near-infrared regions of spectra (e.g., \citealt{allard1997,rojas_ayala2012}). Traditional stellar models lack the necessary line lists/opacity information to reproduce molecular features (e.g., \citealt{mann2013b}), which severely hampers their ability to model M dwarf atmospheres. A few studies have made modifications to line lists and model fitting methodologies that more accurately capture M dwarf chemistry (e.g., \citealt{souto2022,melo2024,hejazi2024}). However, these customized M dwarf modeling pipelines are slow, and have only been used to characterize small ($\sim$20 stars; \citealt{souto2022}) M dwarf samples to date.

This has spurred development of empirical stellar characterization techniques that do not rely on physical stellar models, such as the \emph{The Cannon} \citep{ness2015,casey2016}. A data-driven framework, \emph{The Cannon} uses a training set of spectra from stars with well-determined ``labels" (e.g., stellar parameters and/or elemental abundances) to construct a predictive model of the flux at every pixel in the wavelength range of the spectra. While the training set labels originate from other datasets that may utilize physical stellar models, \emph{The Cannon} is able to construct a model built purely from the labeled training spectra. In that sense, \emph{The Cannon} does not utilize physical stellar models; it only requires a training set of spectra with high quality labels that span the label parameter space of the stars we seek to characterize. The resulting data-driven model can then infer the parameter/abundance labels of other stellar samples given their spectra. This makes \emph{The Cannon} a valuable tool for characterizing cool stars that current stellar models struggle with, e.g., M dwarfs \citep[][]{behmard2019,birky2020,galgano2020,rains2024}. Data-driven approaches are also computationally inexpensive compared to methods that rely on stellar atmosphere modeling, so \emph{The Cannon} is well-suited to characterizing stellar samples from large spectroscopic surveys. 

Here, we use \emph{The Cannon} to infer a wide set of elemental abundances for $\sim$17,000 M dwarfs with Milky Way Mapper (MWM) data from SDSS-V, the current phase of the Sloan Digital Sky Survey. We outline \emph{The Cannon} in Section \ref{sec:the_cannon}, and describe the MWM data and how we processed it for \emph{The Cannon} in Section \ref{sec:data}. We apply \emph{The Cannon} to a small ($\sim$20 stars) benchmark M dwarf sample and evaluate its performance in Section \ref{sec:souto_loocv}. In Section \ref{sec:sample}, we construct a larger M dwarf training set drawn from SDSS-V/MWM and assess its improved performance by testing against the aforementioned benchmark sample and M dwarfs from the Hyades open cluster. We apply the SDSS-V/MWM training set to a sample of $\sim$17,000 SDSS-V/MWM M dwarfs in Section \ref{sec:test_set}, and report their abundances for Fe, Mg, Al, Si, C, N, O, Ca, Ti, Cr, and Ni.

\section{\emph{The Cannon}} \label{sec:the_cannon}

\emph{The Cannon} is a regression model that operates under two key assumptions: that stellar spectra with identical labels look identical at every pixel, and that the flux at every pixel in a spectrum changes continuously as a function of the stellar labels. \emph{The Cannon} can infer labels for stellar samples given their spectra via a two-step process: a ``training" step in which the spectra and labels of stars that compose the training set are used to construct a predictive model of the flux at every pixel in the wavelength range, and a ``test" step in which the model is applied to spectra of other stars in order to infer their labels. In principle, the labels can be any physical parameters that typically parameterize stellar atmosphere models (e.g., $T_{\textrm{eff}}$, log $g$, [Fe/H], etc.), or empirical labels that serve as proxies for physical parameters (e.g., spectral types, colors, magnitudes, etc.) \citep{birky2020}. Whatever the training set labels are, they must be high-quality as the inferred labels will only be as accurate as the training labels, and only precise if the training labels are measured consistently across the training set stars. The training set labels must also span the parameter space of the true labels that we seek to infer for the test set, because \emph{The Cannon} does not extrapolate well outside the training set parameter space. 

In this study, we use \emph{The Cannon 2}, the second implementation of \emph{The Cannon} developed by \citet{casey2016}. Hereafter, we will refer to \emph{The Cannon 2} simply as \emph{The Cannon}. This version allows for more complex flux models than the original, which aids inference of more labels, e.g., a large set of abundances beyond [Fe/H]. It has also been used to successfully infer M dwarf labels in previous studies (e.g., \citealt{behmard2019}). 

\subsection{Training Step}
\emph{The Cannon} flux model for a star $n$ at wavelength pixel $j$ is expressed as

\begin{equation} \label{eq:equation1}
f_{jn} = v(\ell_{n}) \cdot \theta_{j} + e_{jn}, 
\end{equation}

\noindent where $\theta_{j}$ is a vector containing the set of flux model coefficients at each pixel $j$, and $v(\ell_{n})$ is a function of the label list $\ell_{n}$ that is unique for each spectrum $n$. The function $v(\ell_{n})$ is referred to as the ``vectorizer" which can accommodate functions beyond simple polynomial expansions of the label list $\ell_{n}$ (e.g., sums of sines and cosines). The noise term $e_{jn}$ can be taken as drawn from a Gaussian with zero mean and variance $\sigma^{2}_{jn} + s^{2}_{j}$, where $\sigma^{2}_{jn}$ is the flux uncertainty reported for the training set spectra, and $s^{2}_{j}$ is the intrinsic scatter of the model at each pixel $j$. This intrinsic scatter is analogous to the expected deviation of the model from the spectrum at pixel $j$. We can relate the flux model to a single-pixel log-likelihood function:

\begin{equation} \label{eq:equation2}
\begin{split}
\textrm{ln} \hspace{0.5mm} p(f_{jn} | \theta_{j},l_{n},s_{j}^{2}) = - \hspace{0.5mm} \frac{[f_{jn}  \hspace{0.5mm} -  \hspace{0.5mm} v(l_{n}) \cdot \theta_{j}]^{2}}{\sigma_{jn}^{2} + s_{j}^{2}}  \hspace{0.5mm} \\
- \textrm{ln}(\sigma_{jn}^{2} + s_{j}^{2}) + \Lambda_{j}Q(\theta_{j}) \hspace{0.5mm},
\end{split}
\end{equation}

\noindent where $\Lambda_{j}$ is a regularization parameter and $Q(\theta_{j})$ is a regularizing function that encourages the flux model coefficients $\theta_{j}$ to approach zero, which combats overfitting. This is potentially useful for inferring label sets that include many elemental abundances because we expect that only a small number of abundances will affect the flux at each pixel in the wavelength range.

In the training step, each log-likelihood is maximized to derive the best-fit model coefficients $\theta_{j}$ and intrinsic scatter $s^{2}_{j}$ at each wavelength pixel $j$:

\begin{equation} \label{eq:equation3}
\begin{split}
\theta_{j}, s_{j}^{2} = \underset{\textrm{argmax}}{{\theta_{j},s_{j}}}
\Big [\sum_{n=0}^{N-1} \textrm{ln} \hspace{0.5mm} p(f_{jn} | \theta_{j}, l_{n},s_{j}^{2})\Big] \hspace{0.5mm}.
\end{split}
\end{equation}

\subsection{Test Step}
In the test step, we use the optimized model coefficients and scatter ($\theta_{j}$, $s^{2}_{j}$) in our set of single-pixel log-likelihood functions to infer the label list. We do this by maximizing the log-likehood functions, or equivalently, minimizing the negative log-likehood functions:

\begin{equation} \label{eq:equation4}
\begin{split}
l_{n} =  \underset{l_{n}}{\textrm{argmin}} \Big [\sum_{j=0}^{J-1} 
 -\textrm{ln} \hspace{0.5mm} p(f_{jn} | \theta_{j}, l_{n},s_{j}^{2})\Big] \\
= \underset{l_{n}}{\textrm{argmin}} \Big [\sum_{j=0}^{J-1} -  \hspace{0.5mm} \frac{[f_{jn} -  v(l_{n}) \cdot \theta]^{2}}{\sigma_{jn}^{2} + s_{j}^{2}} \Big] \hspace{0.5mm},
\end{split}
\end{equation}

\noindent where we can consider $-\textrm{ln} \hspace{0.5mm} p(f_{jn} | \theta_{j}, l_{n},s_{j}^{2})$ (hereafter $\chi^{2}$) a goodness-of-fit metric that assesses how well the flux model fits the actual spectra of each test set star.

\begin{figure*}[t]
    \centering
    \begin{minipage}{0.99\textwidth}
        \centering
    \includegraphics[width=0.99\textwidth]{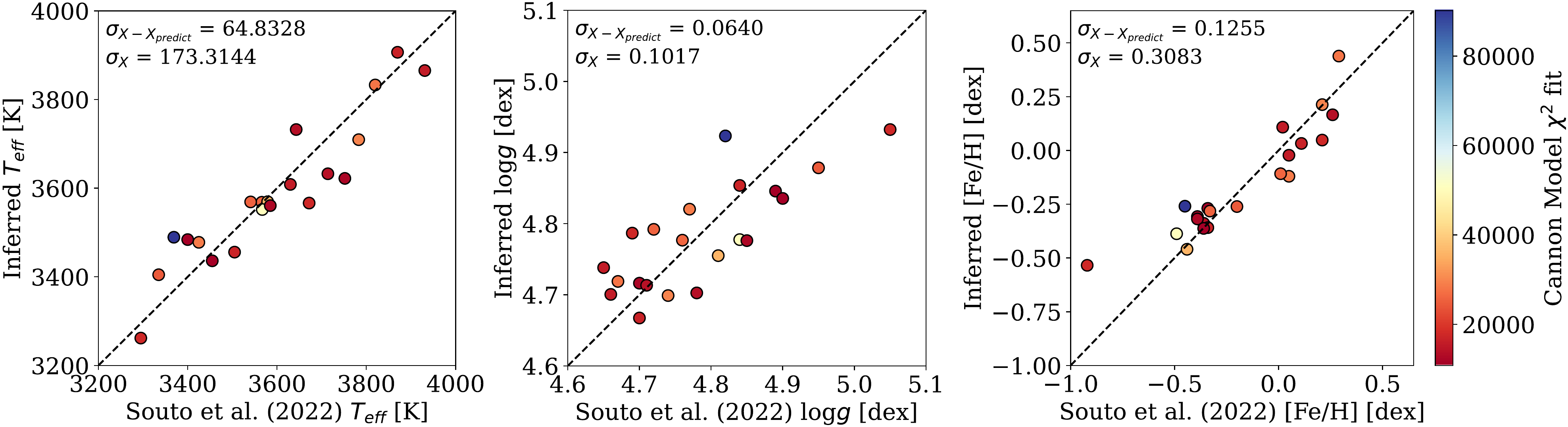} % first figure itself
    \end{minipage} \hfill
    \vspace*{2mm}
    \hspace*{1.05mm}
    \begin{minipage}{0.99\textwidth}
        \centering
        \includegraphics[width=0.98\textwidth]{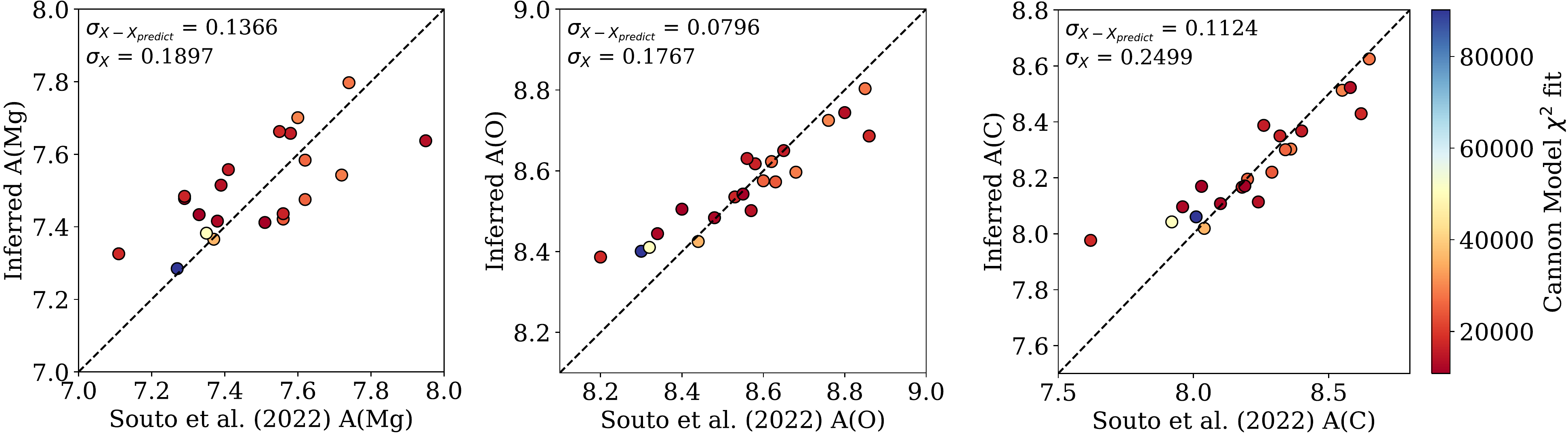} % second figure itself
    \end{minipage} \hfill
    \vspace*{2mm}
    \hspace*{1.6mm}
    \begin{minipage}{0.99\textwidth}
        \centering
        \includegraphics[width=0.98\textwidth]{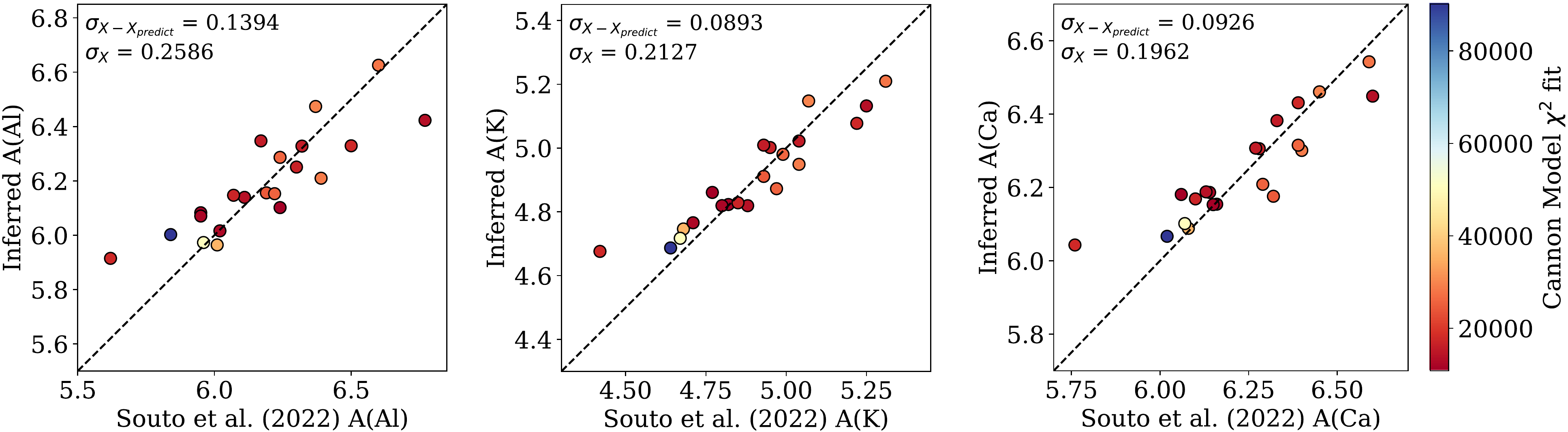} % second figure itself
    \end{minipage} \hfill
    \caption{1-to-1 plots of our inferred versus the reported labels for the \citet{souto2022} sample from our LOOCV scheme with \emph{The Cannon}. The points are colored by the $\chi^{2}$ of the flux model fit. We report the rms scatter between the inferred and \citet{souto2022} labels in the top left of each plot, and the scatter of the \citet{souto2022}-reported values for that label below. The former values are smaller than the latter for every label, indicating that \emph{The Cannon} recovers the labels with rms scatter well within the intrinsic label scatter.}
    \label{fig:figure1}
    \end{figure*}
\newpage

\section{SDSS-V/MWM Data} \label{sec:data}
The SDSS-V/MWM survey has an $H$-band (1.51$-$1.7 $\mu$m) component that employs the Apache Point Observatory Galactic Evolution Experiment (APOGEE) spectrographs, through which it conducts a high resolution ($R$ $\sim$ 22,500) survey that currently provides over 1 million stellar spectra through ongoing observations \citep{majewski2017,wilson2019,almeida2023}. While the APOGEE survey is primarily designed for targeting bright objects, namely red giants, it has surveyed $\sim$50,000 M dwarfs to date. The APOGEE Stellar Parameter and Chemical Abundances Pipeline (ASPCAP) provides abundances for a wide set of elements at typical precisions of $<$0.1 dex \citep{garcia2016}. However, main-sequence stars with $<$4500 K temperatures are known to have problematic ASPCAP abundances, making ASPCAP unreliable for M-dwarfs. This motivates our work on constructing a data-driven model for inferring detailed M dwarf abundances.

APOGEE provides pseudo-continuum-normalized, rest-frame-shifted, co-added spectra across all observed epochs. However, we opt to use the non-pseudo-continuum-normalized spectra because \emph{The Cannon} requires input spectra to be normalized via a linear operation; this is not satisfied by the APOGEE pseudo-continuum-normalization procedure that makes use of high-order polynomials \citep{jonsson2020}. Instead, we carried out continuum normalization via error-weighted, broad Gaussian smoothing with

\begin{equation} \label{eq:equation5}
\begin{split}
\bar{f}(\lambda_{0}) = \frac{\sum_{j} (f_{j} \sigma_{j}^{-2}w_{j}(\lambda_{0}))}{\sum_{j} (\sigma_{j}^{-2} w_{j}(\lambda_{0}))} \hspace{0.5mm},
\end{split}
\end{equation}

\noindent where $f_{j}$ is the flux at pixel $j$ of the wavelength range, $\sigma_{j}$ is the uncertainty at pixel $j$, and the weight $w_{j} (\lambda_{0})$ is drawn from a Gaussian:

\begin{equation} \label{eq:equation6}
\begin{split}
w_{j}(\lambda_{0}) = e^{-\frac{(\lambda_{0} - \lambda_{j})^{2}}{L^{2}}} \hspace{0.5mm} ,
\end{split}
\end{equation}

\noindent where $L$ is chosen to be 10 {{\AA}}, which is slightly larger than typical absorption features in the APOGEE spectra. Gaussian smoothing is often used to continuum normalize spectra in preparation for \emph{The Cannon}, with $L$ adjusted according to the spectral resolution (e.g., \citealt{ho2017,behmard2019,rampalli2021}).

The APOGEE spectra signal-to-noise (SNR) ratios are typically high, with average SNR levels for M dwarfs of approximately 100/pix. However APOGEE M dwarf spectra are often reported with SNR $\gtrsim$ 200/pix. These high SNR targets have underestimated flux uncertainties that artificially inflate flux model spectral fit $\chi^{2}$ values from \emph{The Cannon}. To address this, we set all $<$0.005 flux uncertainty values to 0.005; typical normalized flux uncertainties should scale as $\sim$1/SNR, and APOGEE targets have maximum effective SNR levels of $\sim$200/pix. Thus, flux uncertainty values smaller than 1/200 = 0.005 are unrealistic.

\section{Small Training Set Limitations} \label{sec:souto_loocv}
Previous studies demonstrate that \emph{The Cannon} can infer detailed elemental abundances for red giants \citep{casey2016} and FGK dwarfs (e.g., \citealt{rice2020,rampalli2024,angelo2024}), but it has never been used to infer abundances beyond [Fe/H] or occasionally [Ti/Fe] for M dwarfs \citep{behmard2019,birky2020,rains2024}. This is largely because we lack robust M dwarf training sets with well-determined abundances for other elements. The largest M dwarf dataset with a wide set of abundance measurements is from \citet{souto2022}. This dataset consists of 21 M dwarfs observed during SDSS-IV/APOGEE-1 \citep{blanton2017}, with spectra collected by the APOGEE-N instrument \citep{gunn2006,wilson2019} and reduced as part of the 16th SDSS data release (DR16; \citealt{nidever2015,ahumada2020,jonsson2020}). In short, \citet{souto2022} employed LTE MARCS model atmospheres \citep{gustafsson2008} and the TurboSpectrum spectral synthesis code \citep{alvarez1998,plez2012}, along with a custom line list based on the APOGEE line list from DR17 (e.g., \citealt{smith2013,hasselquist2016,cunha2017}). This enabled them to measure $T_{\textrm{eff}}$, log $g$, Fe, C, O, Mg, Al, K, and Ca for their 21 M dwarfs. 

We find that the small \citet{souto2022} sample is an effective training set for constructing a model with \emph{The Cannon} that reproduces the reported abundances to high precision. We demonstrate this through a leave-one-out cross-validation (LOOCV) scheme in which we train \emph{The Cannon} on all 21 M dwarfs but one, and then infer the detailed abundances of the removed M dwarf with the resultant model trained on $N-$1. We loop through the entire \citet{souto2022} sample with this procedure to infer abundances for each M dwarf. For this test, we use SDSS-V/MWM spectra for the \citet{souto2022} sample rather than the SDSS-IV/APOGEE DR16 spectra used in the original study. The SNR levels range from 60$-$570/pix.

The models are functions of the parameter and abundance labels, listed in the label vector below:

\begin{equation} \label{eq:equation7}
\begin{split}
\ell_{n} = [1,\hspace{1mm} T_{\textrm{eff}},\hspace{1mm} \textrm{log}g,\hspace{1mm} \textrm{[Fe/H]},\hspace{1mm} \textrm{A(Mg)},\hspace{1mm} \textrm{A(O)},\hspace{1mm} \textrm{A(C)},\\ \textrm{A(Al)},\hspace{1mm} \textrm{A(K)},\hspace{1mm} \textrm{A(Ca)}]
\end{split}
\end{equation}

\noindent We test models that are both linear and quadratic functions of the labels. The linear models are significantly faster to train (which is unsurprising given the large set of abundance labels), and exhibit slightly better performance in recovering the reported abundances. This unsurprising because the small \citet{souto2022} sample does not span a large metallicity range, and many elemental abundances are correlated with each other. Thus, quadratic label terms likely introduce excessive/unneeded model complexity. Consequently we employ linear models for this LOOCV test.

The LOOCV results are shown in Figure \ref{fig:figure1}. We find that our LOOCV implementation reproduces the detailed M dwarf abundances reported in \citet{souto2022} to precisions of 0.09$-$0.14 dex. These precisions are affected by the uncertainties on the \citet{souto2022} abundances, which are reported as 0.02$-$0.15 dex and constitute uncertainties on our model labels. To calculate these uncertainties, \citet{souto2022} propagated the uncertainties on their adopted atmospheric parameters ($T_{\textrm{eff}}$, log $g$, [M/H], C/O) and pseudo-continuum normalization procedure \citep{souto2017,souto2022}. While our LOOCV abundance precisions are affected by these reported abundance uncertainties, it is difficult to disentangle their contribution from other error sources in the APOGEE data reduction process, and the inherent scatter of our model. We reserve a longer discussion of uncertainties on our inferred abundances for later in the manuscript (see Section \ref{sec:test_set}). 

The peak flux model spectral fit $\chi^{2}$ of all the \citet{souto2022} M dwarfs is $\sim$12,000, which translates to a reduced $\chi^{2}$ of $\sim$1.6 considering that there are $\sim$7400 pixels in the APOGEE wavelength range (which can be considered the number of degrees of freedom). An excellent flux model fit would yield a reduced $\chi^{2}$ of approximately 1, but our peak $\chi^{2}$ is slightly higher because our flux model fits are imperfect, which is unsurprising given the complexity of M dwarf spectra. Still, a reduced $\chi^{2}$ of $\sim$1.6 indicates a good flux model fit with \emph{The Cannon} (e.g., \citealt{birky2020,rampalli2024}).

There is one M dwarf (SDSS ID = 80419035) in the sample with an anomalously large flux model spectral fit $\chi^{2}$ value of $\sim$90,000. We removed this star and re-ran the LOOCV scheme, and found only marginal improvement to the inferred abundance precisions. This star is a fast rotator with $v$\hspace{0.5mm}sin\hspace{0.5mm}$i$ = 13.5 $\pm$ 2.0 km s$^{-1}$ \citep{souto2020}. Upon visual inspection of its spectrum, we see no obvious rotational broadening, but rapid rotation can induce magnetic activity in late type M dwarfs that affect the flux in other ways (e.g., \citealt{suarez_mascareno2016}). For example, through flares, starspots, and even radii inflation due to magnetic inhibition of convection. All these factors likely contribute to the relatively poor flux fit our model achieves for this star. In general however, this scheme successfully recovers the correct M dwarf abundances to high precisions. This demonstrates, for the first time, that \emph{The Cannon} is capable of inferring precise M dwarf abundances for a wide set of elements.

\begin{deluxetable*}{llrrrrrr}
\setlength{\tabcolsep}{1em}
\tablewidth{0.6\textwidth}
\tabletypesize{\footnotesize}
\tablewidth{0pt}
\tablecaption{M Dwarf Training Set Properties}
\tablecolumns{8}
\tablehead{
\colhead{\emph{Gaia} DR3 ID} &
\colhead{SDSS ID} &
\colhead{$T_{\textrm{eff}}$} &
\colhead{[Fe/H]} &
\colhead{[Mg/H]} &
\colhead{[Al/H]} &
\colhead{[Si/H]} &
\colhead{[C/H] \hspace{5mm}...} \\[-0.2cm]
\colhead{} &
\colhead{} &
\colhead{K} &
\colhead{dex} &
\colhead{dex} &
\colhead{dex} &
\colhead{dex} &
\colhead{\hspace{-8mm}dex} 
}
\startdata
4665710629633988736 & 91725384 & 3910 & $-$0.10 $\pm$ 0.01 & 0.01 $\pm$ 0.01 & 0.01 $\pm$ 0.03 & $-$0.09 $\pm$ 0.02 & $-$0.07 $\pm$ 0.02 \\
4819175927157254784 & 92916465 & 3704 & 0.10 $\pm$ 0.01 & 0.12 $\pm$ 0.02 & 0.12 $\pm$ 0.03 & 0.07 $\pm$ 0.02 & 0.05 $\pm$ 0.03 \\
4763618910272182400 & 92524557 & 3396 & 0.11 $\pm$ 0.01 & 0.14 $\pm$ 0.01 & 0.21 $\pm$ 0.03 & 0.14 $\pm$ 0.02 & 0.14 $\pm$ 0.02 \\
831399531274288512 & 57258526 & 3543 & 0.18 $\pm$ 0.01 & 0.14 $\pm$ 0.02 & 0.25 $\pm$ 0.03 & 0.21 $\pm$ 0.02 & 0.03 $\pm$ 0.03 \\
839567326416583808 & 57301576 & 3382 & 0.06 $\pm$ 0.01 & 0.12 $\pm$ 0.02 & 0.05 $\pm$ 0.03 & 0.02 $\pm$ 0.02 & 0.08 $\pm$ 0.02 \\
3922083599776523520 & 80399914 & 3637 & $-$0.10 $\pm$ 0.01 & $-$0.10 $\pm$ 0.02 & $-$0.15 $\pm$ 0.03 & $-$0.13 $\pm$ 0.02 & $-$0.13 $\pm$ 0.03 \\
1534086421765263360 & 61843665 & 3654 & $-$0.29 $\pm$ 0.01 & $-$0.03 $\pm$ 0.01 & $-$0.06 $\pm$ 0.03 & $-$0.15 $\pm$ 0.02 & $-$0.09 $\pm$ 0.02 \\
2581806619466249728 & 70532624 & 3729 & $-$0.26 $\pm$ 0.01 & -0.15 $\pm$ 0.01 & $-$0.20 $\pm$ 0.03 & $-$0.20 $\pm$ 0.02 & $-$0.31 $\pm$ 0.03 \\
315289980082496256 & 116701844 & 3551 & $-$0.08 $\pm$ 0.01 & $-$0.14 $\pm$ 0.01 & $-$0.17 $\pm$ 0.03 & $-$0.15 $\pm$ 0.02 & $-$0.15 $\pm$ 0.02 \\
3637535866023121536 & 78881207 & 3735 & 0.08 $\pm$ 0.01 & 0.12 $\pm$ 0.01 & 0.13 $\pm$ 0.03 & 0.10 $\pm$ 0.02 & 0.09 $\pm$ 0.02 \\
935566511271187072 & 58005724 & 3419 & 0.22 $\pm$ 0.01 & 0.26 $\pm$ 0.01 & 0.33 $\pm$ 0.03 & 0.25 $\pm$ 0.02 & 0.27 $\pm$ 0.02 \\
678053878665346816 & 56426757 & 3337 & 0.01 $\pm$ 0.01 & $-$0.04 $\pm$ 0.01 & $-$0.10 $\pm$ 0.03 & $-$0.03 $\pm$ 0.02 & 0.02 $\pm$ 0.02 \\
702082159096764288 & 56575737 & 3516 & 0.03 $\pm$ 0.01 & 0.02 $\pm$ 0.01 & 0.00 $\pm$ 0.03 & 0.03 $\pm$ 0.02 & $-$0.01 $\pm$ 0.02 \\
& & & ...\\
\enddata
\tablecomments{This table lists the properties of the M dwarfs in our FGK-M training set drawn from SDSS-V/MWM. The properties consist of all labels used to construct our flux models with \emph{The Cannon} (photometric temperatures and abundances for all elements of interest). The abundances and their errors are the reported ASPCAP abundances of the FGK companions. We only list a subset of the abundances in this table, but the full set is provided in the downloadable version.
\vspace{1.5mm}
\newline(This table is available in its entirety in machine-readable form.)}
\end{deluxetable*} \label{tab:table1}

\section{FGK-M Training Set} \label{sec:sample}
We use SDSS-V/MWM to construct a larger training set of 79 M dwarfs (compared to 21 M dwarfs in the \citet{souto2022} sample) with detailed abundance measurements. Because ASPCAP elemental abundances are unreliable for M dwarfs, we cannot simply use the ASPCAP abundances as labels for \emph{The Cannon}. Instead, we tag M dwarfs with the ASPCAP abundances of FGK dwarf binary companions. Unlike M dwarfs, FGK stars are well-modeled by ASPCAP because physical stellar models and synthetic spectra are reliable in the solar-like regime. Tagging M dwarfs with the abundances of solar-like binary companions is considered a ``gold standard" method (e.g., \citealt{mann2013a,mann2014,newton2014,maldonado2020,duque_arribas2024}). It is based on the assumption that binary companions share a parent molecular cloud and thus formed from the same material, making them approximately chemically homogeneous at birth (e.g., \citealt{de_silva2007,de_silva2009,bland_hawthorn2010}). However, there is a caveat to this assumption; solar-like stars are affected by diffusion processes (i.e., gravitational settling and radiative levitation) that alter their surface abundances over time \citep{dotter2017,souto2019}. Conversely, M dwarfs are relatively immune from diffusion effects because they have deeper convective envelopes, and diffusion efficiency decreases with depth (e.g., \citealt{liu2019,moedas2022,wanderley2023}). This may produce abundance variations of 0.01$-$0.12 dex between FGK and M dwarf companions assuming solar age \citep{choi2016}. We can potentially correct for diffusion processes by reporting abundances in the form of [X/Fe] rather than [X/H] because diffusion effects are roughly similar across different elements. This was done by \citet{souto2022} to enable comparison of their M dwarf abundances to FGK companion abundances. We leave this as an option in our analysis if diffusion appears to be an issue.

\begin{figure*}[t]
    \centering
    \hspace*{1.15mm}
    \begin{minipage}{0.98\textwidth}
        \centering
    \includegraphics[width=0.99\textwidth]{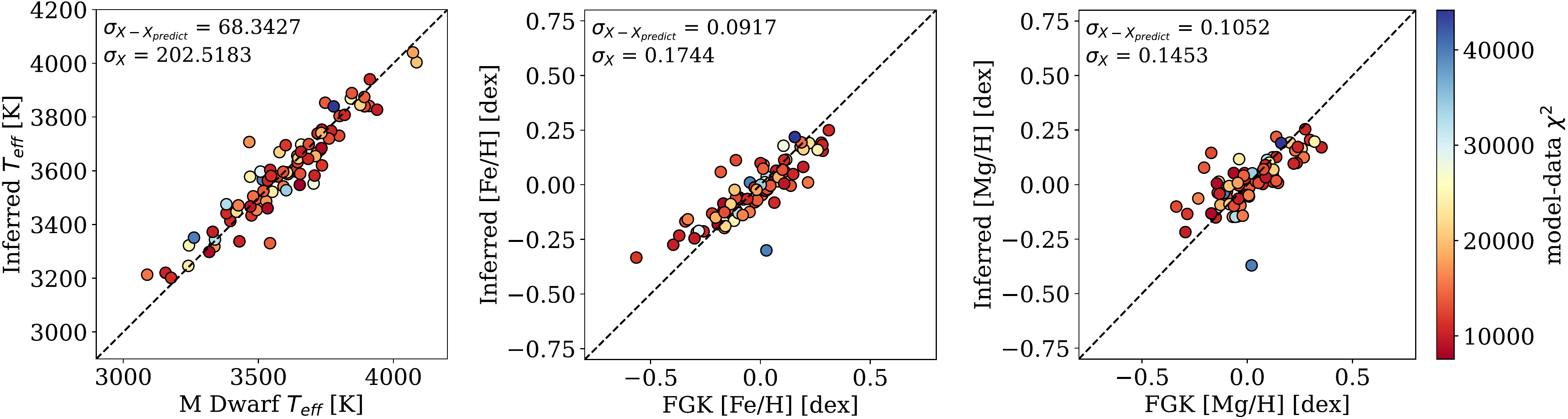} % first figure itself
    \vspace*{1mm}
    \end{minipage} \hfill
    
    \begin{minipage}{0.99\textwidth}
        \centering
        \includegraphics[width=0.988\textwidth]{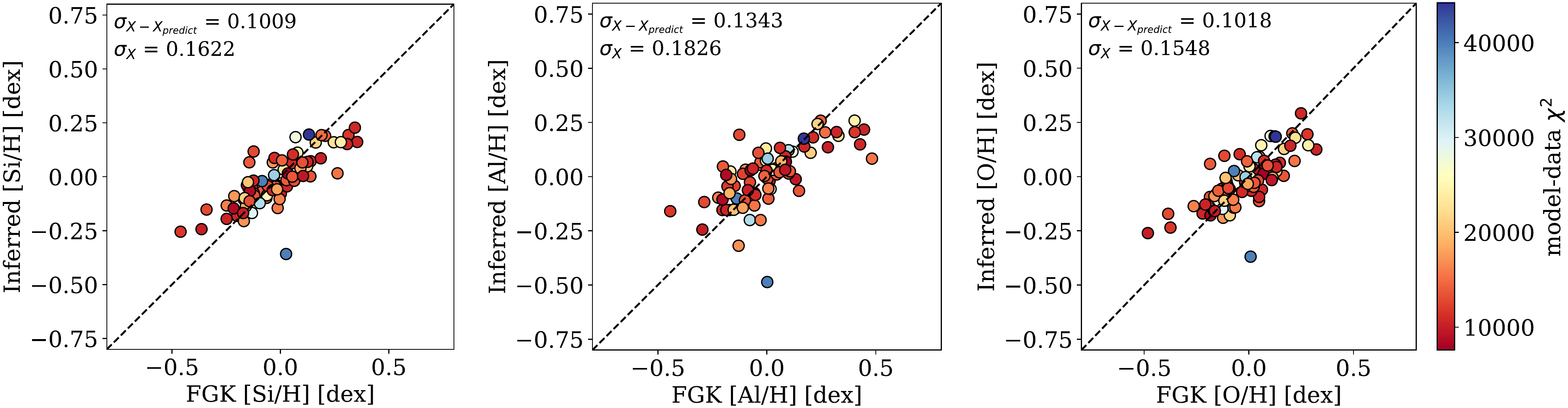} % second figure itself
    \end{minipage} \hfill
    \vspace*{1mm}
    %\hspace*{0.1mm}
    \begin{minipage}{0.999\textwidth}
        \centering
        \includegraphics[width=0.98\textwidth]{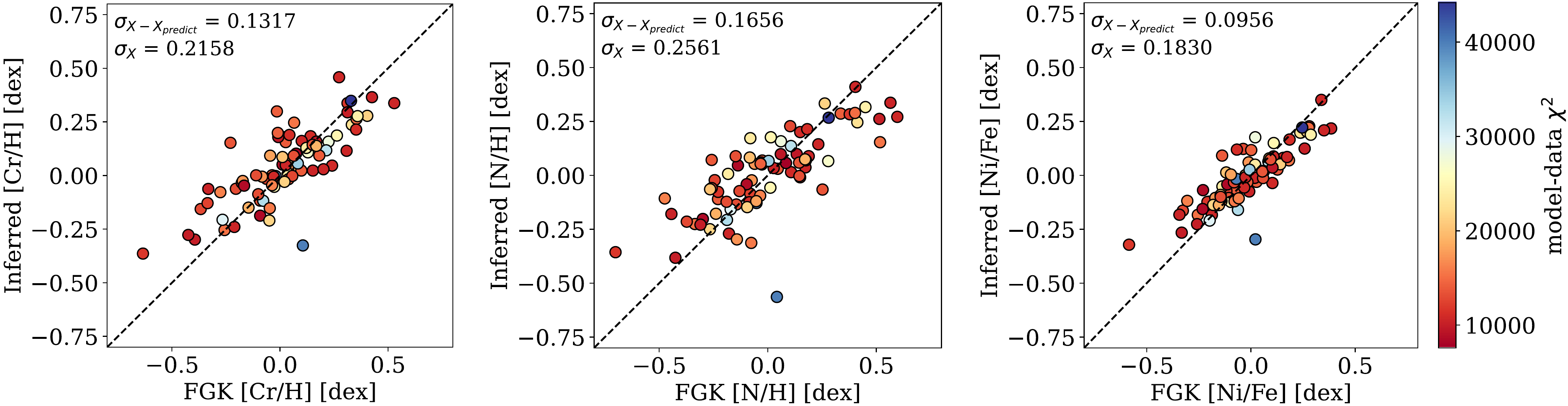} % second figure itself
    \end{minipage} \hfill
    \vspace*{1mm}
    \begin{minipage}{0.999\textwidth}
        \centering
        \includegraphics[width=0.98\textwidth]{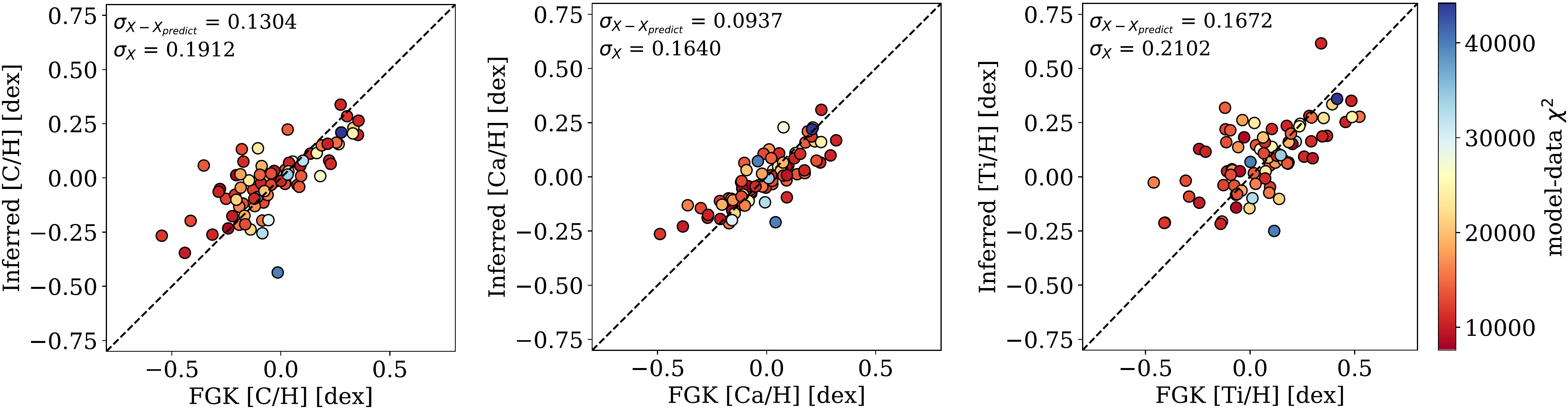} % second figure itself
    \end{minipage} \hfill
    
    \caption{1-to-1 plots of our inferred versus the reported labels for our FGK-M training set after applying LOOCV with \emph{The Cannon}. The points are colored by the $\chi^{2}$ of the flux model spectral fit. We report the rms scatter between the inferred M dwarf and FGK companion labels in the top left of each plot, and the scatter of the FGK companion labels below. The former values are smaller than the latter for every label, indicating that \emph{The Cannon} recovers the labels using the M dwarf spectra with rms scatter that are well within the intrinsic FGK label scatter.}
    \label{fig:figure2}
    \end{figure*}
\newpage

To identify FGK-M binary systems in SDSS-V/MWM, we cross-match all SDSS-V stars with ASPCAP results with the \citet{elbadry2021} binary catalog. Because the ASPCAP log $g$ values are unreliable for cool M dwarfs, we select M dwarf secondaries in binary systems according to M dwarf type-color sequence relations from \citet{pecaut2013ApJS}:

\begin{equation}\label{eq:equation8}
\begin{split}
G - RP > 0.92 \\
G + 5\textrm{log}(\varpi/100) > 8.16
\end{split}
\end{equation}

\noindent where $G$ and $RP$ are photometric passbands and $\varpi$ is the parallax, all from \emph{Gaia} DR3 \citep{gaiadr3}. To ensure the primaries in the remaining binary systems are FGK dwarfs with reliable abundances, we apply the following ASPCAP parameter and abundance cuts:

\begin{equation}\label{eq:equation9}
\begin{split}
\textrm{log}g > 4 \hspace{1mm} \textrm{dex} \\
T_{\textrm{eff}} = 4500-6500 \hspace{1mm} \textrm{K} \\
[\textrm{X/H}]_{\textrm{err}} < 0.2 \hspace{1mm} \textrm{dex} \\
\textrm{Flags} \hspace{1mm} \texttt{x\_h\_flags} \hspace{1mm} \textrm{not set} 
\end{split}
\end{equation}

The $T_{\textrm{eff}}$ cut delineates the range within which ASPCAP elemental abundances are reliable. Finally, we apply a cut of SNR $>$ 50 pix$^{-1}$ to the M dwarfs to ensure that their spectra are sufficiently high quality for training robust flux models with \emph{The Cannon}. This resulted in a final training set of 79 FGK-M binaries with SNR = 51$-$440/pix. The FGK companions span $-$0.56 $<$ [Fe/H] $<$ 0.31 dex, and the abundance uncertainties across all elements we use as training labels range from 0.01$-$0.14 dex. 

We provide the M dwarf training set labels in Table \ref{tab:table1}. These consist of the M dwarf $T_{\textrm{eff}}$ values, and elemental abundances from the FGK companions. While the ASPCAP $T_{\textrm{eff}}$ values are not bad, they are still derived from extrapolations to models that are not designed to accommodate M dwarfs. For this reason, it is hard to say how truly reliable they are across the entire cool temperature range (3000$-$4000 K) spanned by our \emph{Cannon} model. A handful of our training set M dwarfs also lack ASPCAP $T_{\textrm{eff}}$. Instead, we use an empirical color-temperature relation \citep{curtis2020} and \emph{Gaia} DR3 photometry to calculate new $T_{\textrm{eff}}$ values as training set labels. We check the color extinction values using the \texttt{Bayestar19} 3D dust map \citep{Green2019} implemented in the \texttt{dustmaps} Python package \citep{Green2018}. The extinction coefficients are taken from \citet{danielski2018}. $>$95\% of our training set M dwarfs have E(BP$-$RP) $<$ 0.03, corresponding to potential $T_{\textrm{eff}}$ shifts due to reddening of $<$20 K. Two of our M dwarfs have higher extinction values of $\sim$0.06 and $\sim$0.16, corresponding to potential $T_{\textrm{eff}}$ shifts of $\sim$40 K and $\sim$140 K, but both these stars have reported ASPCAP $T_{\textrm{eff}}$ that agree to within 60 K of their photometric $T_{\textrm{eff}}$ from the \citet{curtis2020} relation. We conclude that the photometric $T_{\textrm{eff}}$ values of our training set M dwarfs are not affected by reddening.

After training our model on this FGK-M binary sample, we provide the flux model coefficients $\theta_{j}$ at each pixel $j$ for each label in Figures \ref{fig:figureA1}, \ref{fig:figureA2}, and \ref{fig:figureA3}. We compare the abundance label coefficients with the APOGEE line list \citep{smith2021}, and additional M dwarf lines identified in \citet{souto2022}. We find that the coefficient amplitudes are often large at the locations of strong absorption features. This indicates that our flux models include reliable abundance information.

\begin{figure*}[t]
    \centering
        \includegraphics[width=0.85\textwidth]{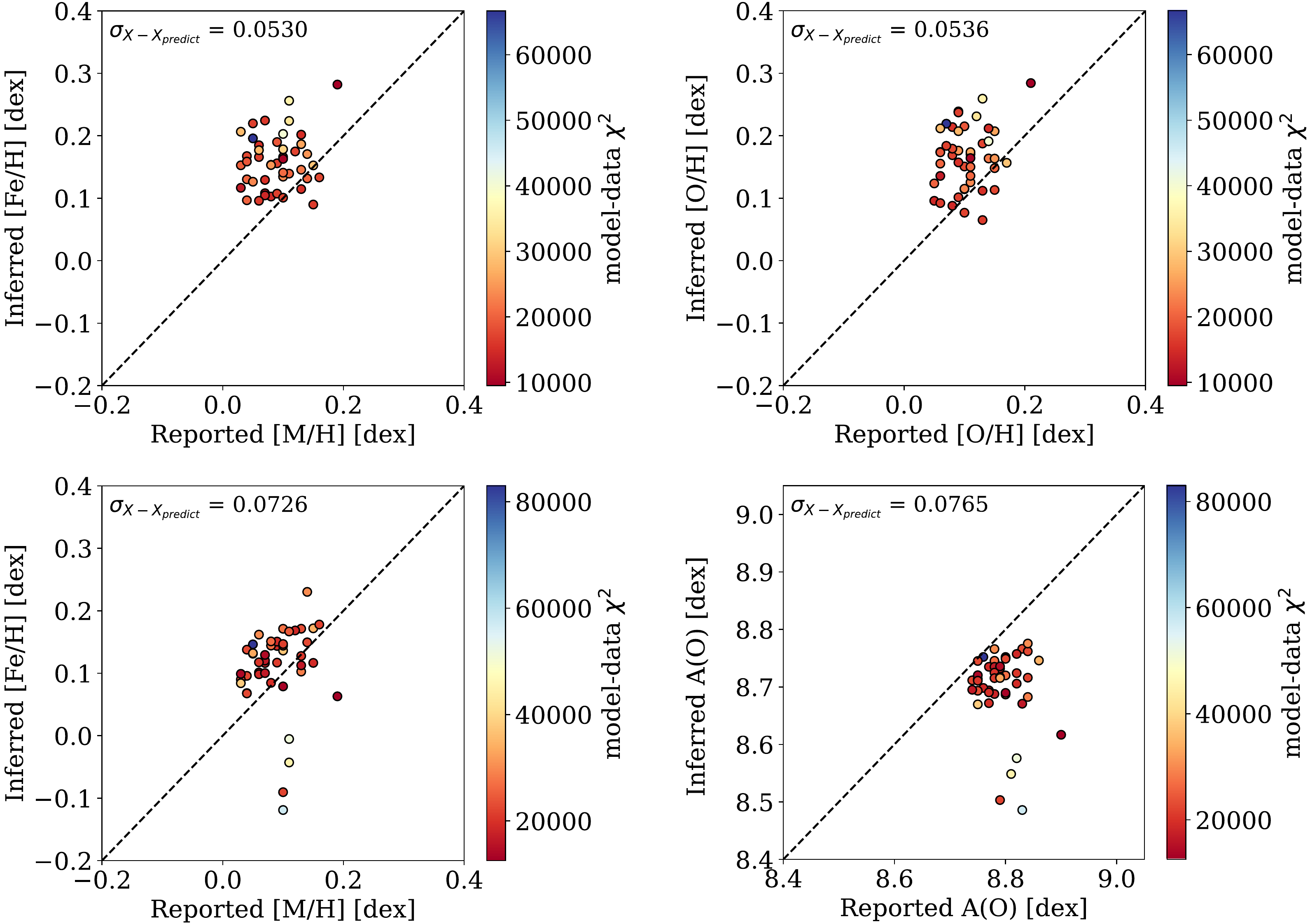} % second figure itself
    \caption{1-to-1 plots of our inferred abundances vs. the reported abundances of M dwarfs from the Hyades cluster using either our FGK-M sample (top row) or the \citet{souto2022} sample (bottom row) as the training set. The points are colored by the $\chi^{2}$ of the flux model spectral fit. We report the rms scatter between the inferred and reported abundance values from \citet{wanderley2023} in the top left of each plot, and the scatter of the reported values for that label below. The rms scatter from training on our FGK-M sample ($\sim$0.05 dex) are lower than those from training on the \citet{souto2022} sample ($\sim$0.07$-$0.08 dex).}
\label{fig:figure3}
\end{figure*}

\subsection{Leave-One-Out Cross-Validation} \label{sec:sdssv_loocv}
To assess the validity of the FGK-M abundance labels, we apply LOOCV cross-validation. Our label vector again corresponds to a linear model, justified by the narrow metallicity range of our FGK-M training set. It includes a large set of abundances:

\begin{equation} \label{eq:equation10}
\begin{split}
\ell_{n} = [1,\hspace{1mm} T_{\textrm{eff}},\hspace{1mm}
\textrm{[Fe/H]},\hspace{1mm} \textrm{[Mg/H]},\hspace{1mm} \textrm{[Al/H]},\hspace{1mm} \textrm{[Si/H]},\\ \textrm{[C/H]},\hspace{1mm} \textrm{[N/H]},\hspace{1mm} \textrm{[O/H]},
\textrm{[Ca/H]},\hspace{1mm} \textrm{[Ti/H]},\hspace{1mm} \\
\textrm{[V/H]},\hspace{1mm}\textrm{[Cr/H]},\hspace{1mm} \textrm{[Ni/H]} ]
\end{split}
\end{equation}

The elemental abundance labels are a combination of those considered most reliable for dwarfs from the ASPCAP pipeline (Fe, C, Mg, Si, Ni), and those involved in common molecular species in M dwarf atmospheres (O, N, Ti, V, Ca, Cr) (e.g., \citealt{rajpurohit2018}). We also include Al because it is well-represented in bulk Earth composition \citep{mcdonough2003}, and is therefore common in rocky planet-forming material. Including this wide set of abundances in our model results in optimal performance; subsets of these abundances do not recover the known training set labels as well. This indicates that each of these elements contain spectral information content, and contribute significantly to M dwarf chemical compositions. The M dwarf $T_{\textrm{eff}}$ label values are derived from the \citet{curtis2020} color-temperature relation as outlined earlier. We do not include log $g$ as a label in our model because M dwarfs evolve slowly, so their physical properties do not change much over the age of the universe after they reach the zero-age main sequence. Because of this, the information contained in M dwarf log $g$ and metallicity can be considered redundant \citep{birky2020}.

We illustrate our LOOCV results in Figure \ref{fig:figure2}. The $\chi^{2}$ values for all M dwarfs in our training set peak at $\sim$10,000 (reduced $\chi^{2}$ of $\sim$1.4), indicating that they are well-fit by the flux models from \emph{The Cannon}. There is one noticeable outlier star in the 1-to-1 abundance plots with a relatively high $\chi^{2}$. It does not have an unusual SNR, FGK companion separation, or reddening level according BP$-$RP color. It also seems to occupy a well-populated area of abundance parameter space (solar-like) according to the abundances of its FGK companion. Still, it is possible that this star is not well-represented by the $N$-1 training set stars in terms of its parameters and/or chemistry. It also does not appear to be a fast rotator based on visual inspection of the flux model fits, but it may still be rapidly rotating as in the case of SDSS ID = 80419035 in the \citet{souto2022} sample. Thus, its relatively poor flux fit may be due to rapid rotation and/or magnetic activity.

In terms of precision, LOOCV recovers the photometric M dwarf $T_{\textrm{eff}}$ to 68 K, and the FGK companion abundance labels to 0.09$-$0.17 dex in rms scatter. These values are smaller than the scatter of the FGK labels themselves (second row of values in top left corners of each Figure \ref{fig:figure2} panel), indicating that our inferred labels are high precision, and are not just reproducing the input label scatter. The only exception is vanadium, which has a high rms scatter of 0.33 dex and does not exhibit a convincing 1-to-1 trend between the inferred and FGK abundances. For this reason we do not include V in Figure \ref{fig:figure2}. This suggests that the ASPCAP V abundances for the FGK dwarf companions are inaccurate, or are not mapped well to the spectra of their M dwarf companions. The former may be true considering that V is measured from a single, weak line in APOGEE spectra (e.g., \citealt{grilo2024}), and the latter may also be true because hyperfine structure affects vanadium spectral features of cool versus solar-like stars differently \citep{shan2021}. Nevertheless, including V as a label in our model improves the precision of our other inferred abundances, indicating that the V abundances still contain useful information. 

We also ran our LOOCV scheme with FGK abundance labels in the form of [X/Fe] rather than [X/H], but find that our results are worse in terms of inferred abundance precisions and 1-to-1 trends. We conclude that there is no noticeable advantage to using [X/Fe] in the interest of mitigating differential diffusion effects on the surface abundances of FGK and M dwarf companions. It is possible that any advantage is wiped out by compounding the uncertainties on Fe and X, where X is any other element of interest. Additionally, X and Fe will be strongly correlated if they share nucleosynthetic channels, which will cause [X/Fe] to lack much of the information inherent in X. In such cases, abundances in the form [X/Fe] will not be informative labels for training \emph{The Cannon}. We conclude that [X/H] abundance labels are best for our purposes.

As in the case of LOOCV on the \citet{souto2022} sample, our abundance precisions from LOOCV on our FGK-M sample are affected by the label uncertainties. In this case, these could be considered the reported uncertainties on the FGK companion ASPCAP abundances. However, it is not clear how reliable the ASPCAP abundance uncertainties are (discussion with A. R. Casey). In our final sample, we use M dwarfs with repeat APOGEE observations to more robustly determine uncertainties on our inferred abundances (see Section \ref{sec:test_set}).

\begin{figure}[t]
    \centering
    \hspace{-5mm}
        \includegraphics[width=0.45\textwidth]{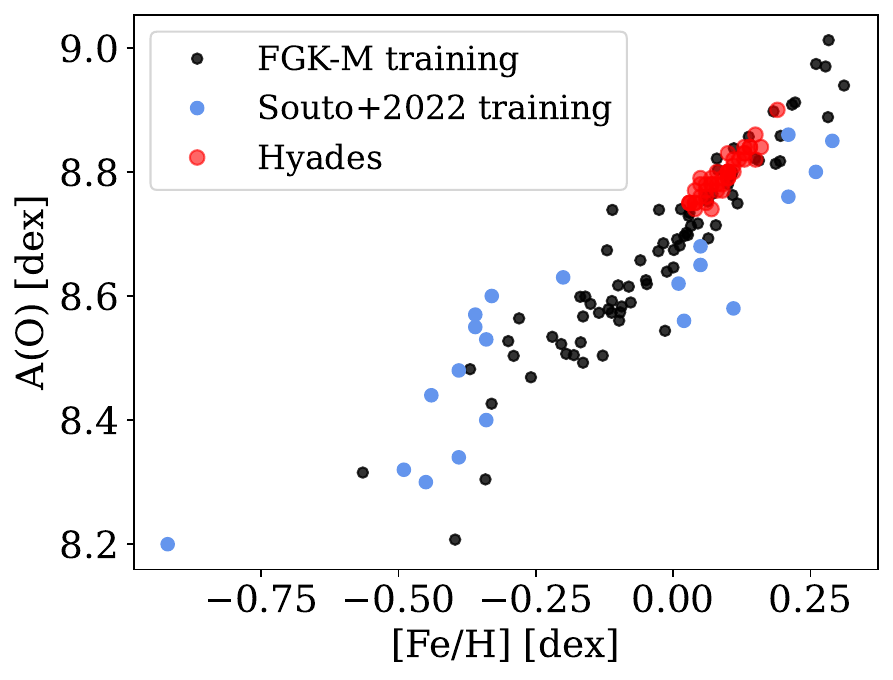} % second figure itself
    \caption{The A(O) and [M/H]/[Fe/H] parameter space spanned by the Hyades M dwarfs from \citet{wanderley2023} (red), the training set of 21 \citet{souto2022} M dwarfs (blue), and our FGK-M training set (black). Because the \citet{souto2022} sample is small, its abundance parameter space is small and sparse, and does not cover the Hyades sample. The region of [M/H]/[Fe/H] $\approx$ 0.1 dex and A(O) $\approx$ 8.8 dex is particularly sparse, and likely responsible for the handful of large outliers in inferred vs. reported abundances in the lower panels of Figure \ref{fig:figure3}.}
\label{fig:figure4}
\end{figure}

\begin{figure*}[t]
    \centering
    \includegraphics[width=0.988\textwidth]{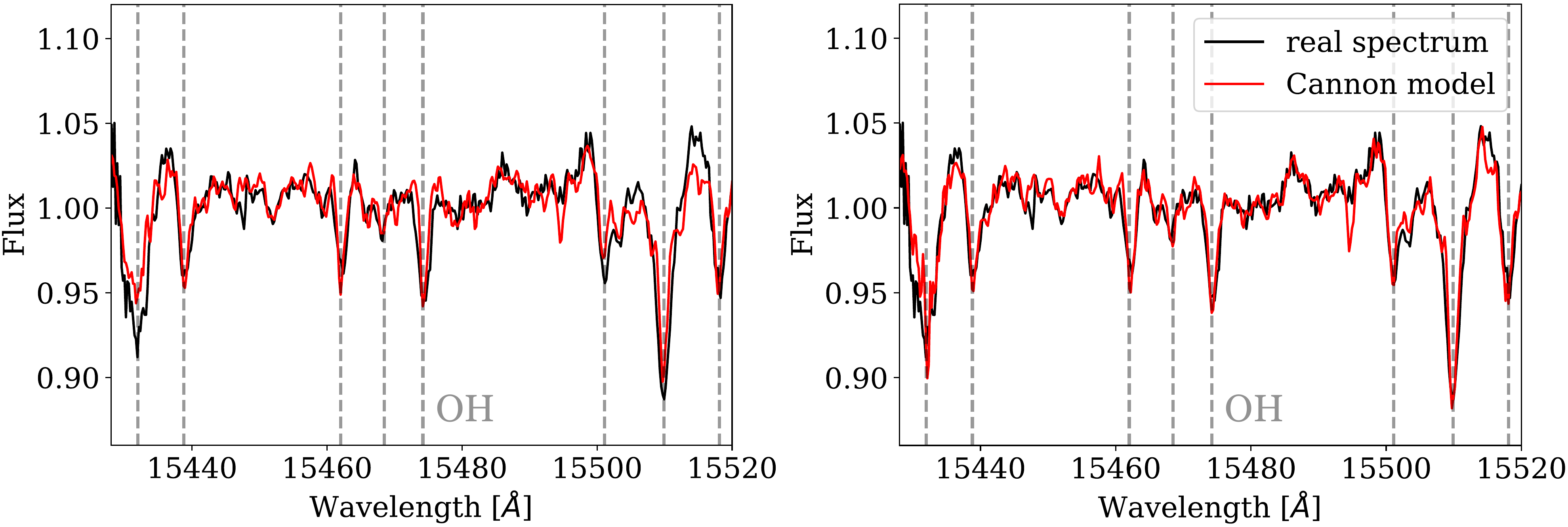} % second figure itself
\caption{Plots of flux model fits to the spectrum of an example M dwarf from the Hyades cluster (SDSS ID = 77382206). The left panel displays the fit resulting from training on the \citet{souto2022} sample, and the right panel displays the fit from training on the FGK-M sample. In both panels the flux model fit is in red, and the real spectrum is in black. Prominent OH lines used for calculating [O/H] or A(O) are marked by the dashed gray lines. It is apparent that the model fit resulting from the FGK-M training set is a better fit to the M dwarf spectrum (right panel).}
\label{fig:figure5}
\end{figure*}

\begin{figure*}[t]
    \centering
    \includegraphics[width=0.988\textwidth]{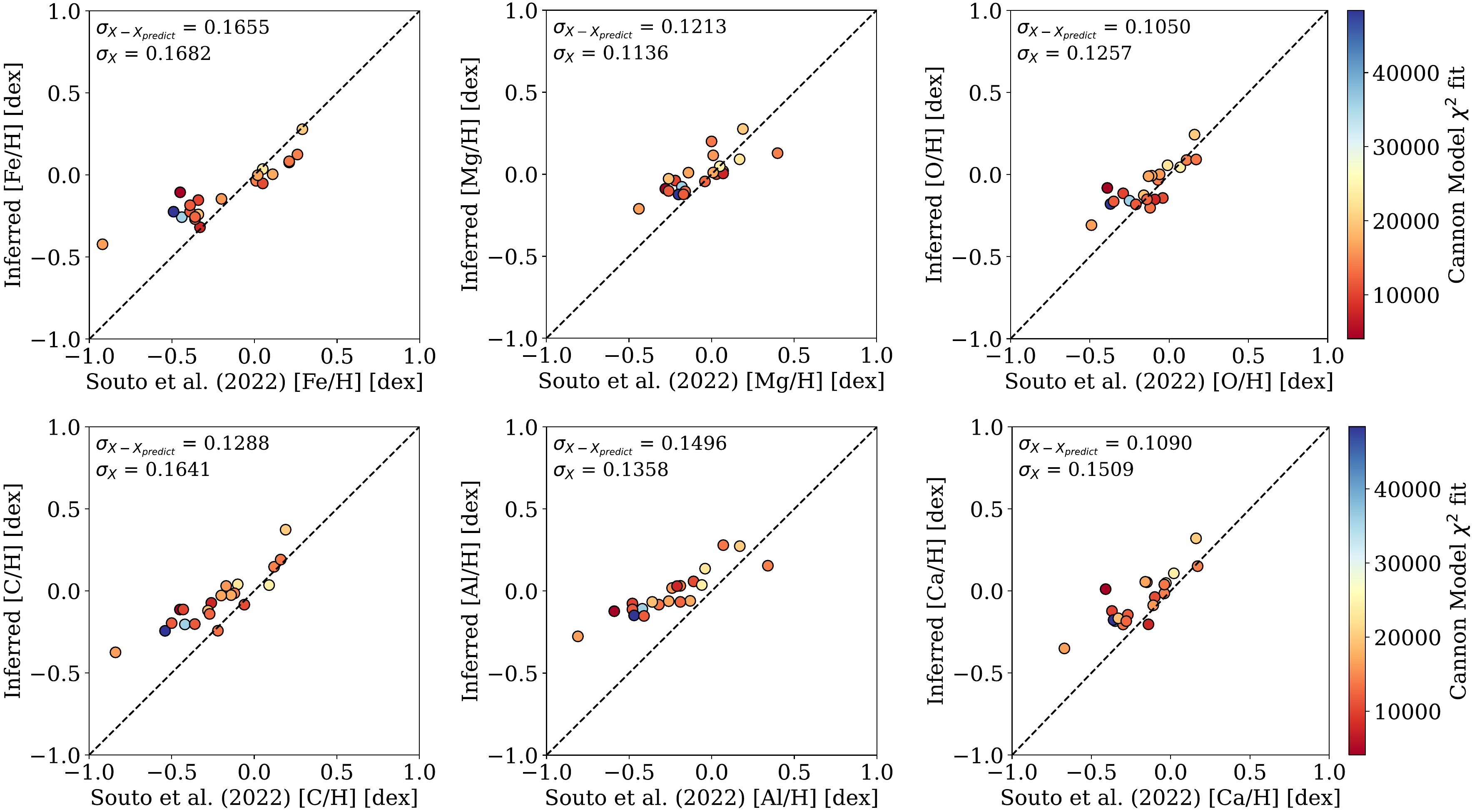} % second figure itself
\caption{1-to-1 plots of the inferred versus reported abundances for M dwarfs from \citet{souto2022} using our FGK-M training set. The points are colored by the $\chi^{2}$ of the flux model spectral fits. We report the rms scatter between the inferred and reported abundance values from \citet{wanderley2023} in the top left of each plot, and the scatter of the reported values for that label below.}
\label{fig:figure6}
\end{figure*}
\newpage

\subsection{Validation with Hyades and Souto et al. (2022) Datasets} \label{sec:validation}

To examine how much better our FGK-M sample performs as a training set compared to the \citet{souto2022} sample, we tested how well each reproduces reported [M/H] and A(O) measurements for M dwarfs in the Hyades open cluster \citep{wanderley2023}. We use the same implementation of \emph{The Cannon} as in our FGK-M LOOCV test, namely a linear model in $T_{\textrm{eff}}$ and all considered elemental abundances in Equation \ref{eq:equation10}. Our results are shown in Figure \ref{fig:figure3}. Our FGK-M sample (Figure \ref{fig:figure3}, upper row) does a noticeably better job at reproducing the Hyades abundances from \citet{wanderley2023} compared to the \citet{souto2022} sample (Figure \ref{fig:figure3}, lower row). In the latter case there are a handful of M dwarfs with very discrepant inferred versus reported abundances. This is likely because these few Hyades M dwarfs fall into a sparsely populated region of the abundance parameter space spanned by the \citet{souto2022} sample, which is small and poorly populated in general because the \citet{souto2022} sample is small (Figure \ref{fig:figure4}, blue points). In contrast, training on our FGK-M sample does not result in large discrepancies between inferred versus reported Hyades M dwarf abundances. Our FGK-M sample performs better because it is bigger than the \citet{souto2022} sample, and thus spans a larger and more well-populated abundance parameter space (Figure \ref{fig:figure4}, black points). Our inferred abundances agree with the reported Hyades abundances to within $\sim$0.05 dex (as opposed to $\sim$0.07$-$0.08 dex with the \citet{souto2022} training set). For reference, the scatter of the metallicities and oxygen abundances reported in \citet{wanderley2023} is 0.03 dex. Compared to these reported abundances, our inferred abundances also adhere more convincingly to the 1-to-1 trend. We also do not see any significant inferred abundance trends with $T_{\textrm{eff}}$, which is reassuring. The flux model spectral fit $\chi^{2}$ values are also lower compared to those from training on \citet{souto2022}, which peak at $\sim$16,000 as opposed to $\sim$23,000 when we trained on the \citet{souto2022} sample. Figure \ref{fig:figure5} illustrates the improved flux model fit for an example Hyades M dwarf from the \citet{wanderley2023} sample, whose abundances are not well-inferred using the \citet{souto2022} training set. The left panel exhibits the model fit from \citet{souto2022}, and the right panel exhibits the model fit from our larger FGK-M sample. The latter fit is a clear improvement, especially around prominent OH lines used to infer the oxygen abundance [O/H] or A(O).
%A $\chi^{2}$ $<$ 100,000 cut is motivated by \citet{birky2020}, who used it to select M dwarfs with spectra well-fit by a \emph{Cannon} model. 

We use the Hyades sample to test adding regularization to our model. A regularization parameter $\Lambda$ is built into \emph{The Cannon} and can be assigned different strengths, which encourage model coefficients to take on zero values. This results in simpler models that are less prone to overfitting. We test regularization parameter values ranging from $\Lambda$ = 10$^{2}$ to 10$^{5}$, and find that including regularization always results in less precise label predictions. For this reason we do not include regularization in our model. This is somewhat surprising as \citet{casey2016} found that including regularization resulted in better inference of a wide set of red giant elemental abundances. However, it is possible that many different abundances contribute to the flux at each wavelength point in M dwarf spectra, making overfitting less of an issue. This would make sense given the complexity of M dwarf spectra (e.g., large number of molecular lines) due to the low $T_{\textrm{eff}}$ of M dwarf atmospheres. 

We also assess how well our FGK-M training set recovers the reported abundances of the \citet{souto2022} sample. Our training label set overlaps with the \citet{souto2022} abundances for Fe, Mg, Al, C, O, and Ca. We again use the same implementation of \emph{The Cannon} as in our LOOCV test and train on all labels in Equation \ref{eq:equation10}. Our inferred versus reported abundances are presented in Figure \ref{fig:figure6}. We do not show our 1-to-1 plot for $T_{\textrm{eff}}$, but the inferred and reported values agree to within 48 K and exhibit a strong 1-to-1 trend. The abundances agree to within 0.1$-$0.17 dex, with good adherence to the 1-to-1 trends. However, our model does struggle to reproduce the low metallicities of \citet{souto2022} stars outside the range of our training set ([Fe/H] $<$ $-$0.56 dex), namely the lowest metallicity \citet{souto2022} M dwarf ([Fe/H] = $-$0.92 dex). We would ideally exclude such low metallicity sources from our analysis of SDSS-V/MWM M dwarfs, but we cannot cut on the ASPCAP [Fe/H] values because they are unreliable. The lowest metallicity \citet{souto2022} star also does not have a high model fit $\chi^{2}$ value, or noticeably poor inferences for other labels (e.g., $T_{\textrm{eff}}$) that could compensate to produce a good model fit. We conclude that there is no clear way to identify M dwarfs with metallicities outside the range of our training set parameter space beforehand. However, we do not expect many SDSS-V/MWM M dwarfs to have metallicities outside $-$0.56 $<$ [Fe/H] $<$ 0.31 dex. Because M dwarfs are faint, those observed will be mostly within the solar neighborhood, and the majority of solar neighborhood main sequence dwarfs are within this metallicity threshold (e.g., \citealt{bensby2014}). Still, we highlight this to caution the reader, and encourage users of our M dwarf model to inspect the flux model fits in cases where the M dwarf metallicity is suspected to be outside this range. Overall though, we reproduce the reported \citet{souto2022} abundances and Hyades [M/H] and A(O) abundances from \citet{wanderley2023} well, demonstrating that our M dwarf model trained on our sample of FGK-M binaries is robust.

\begin{figure}[t]
\hspace*{-0.3cm}
    \centering
    \includegraphics[width=0.48\textwidth]{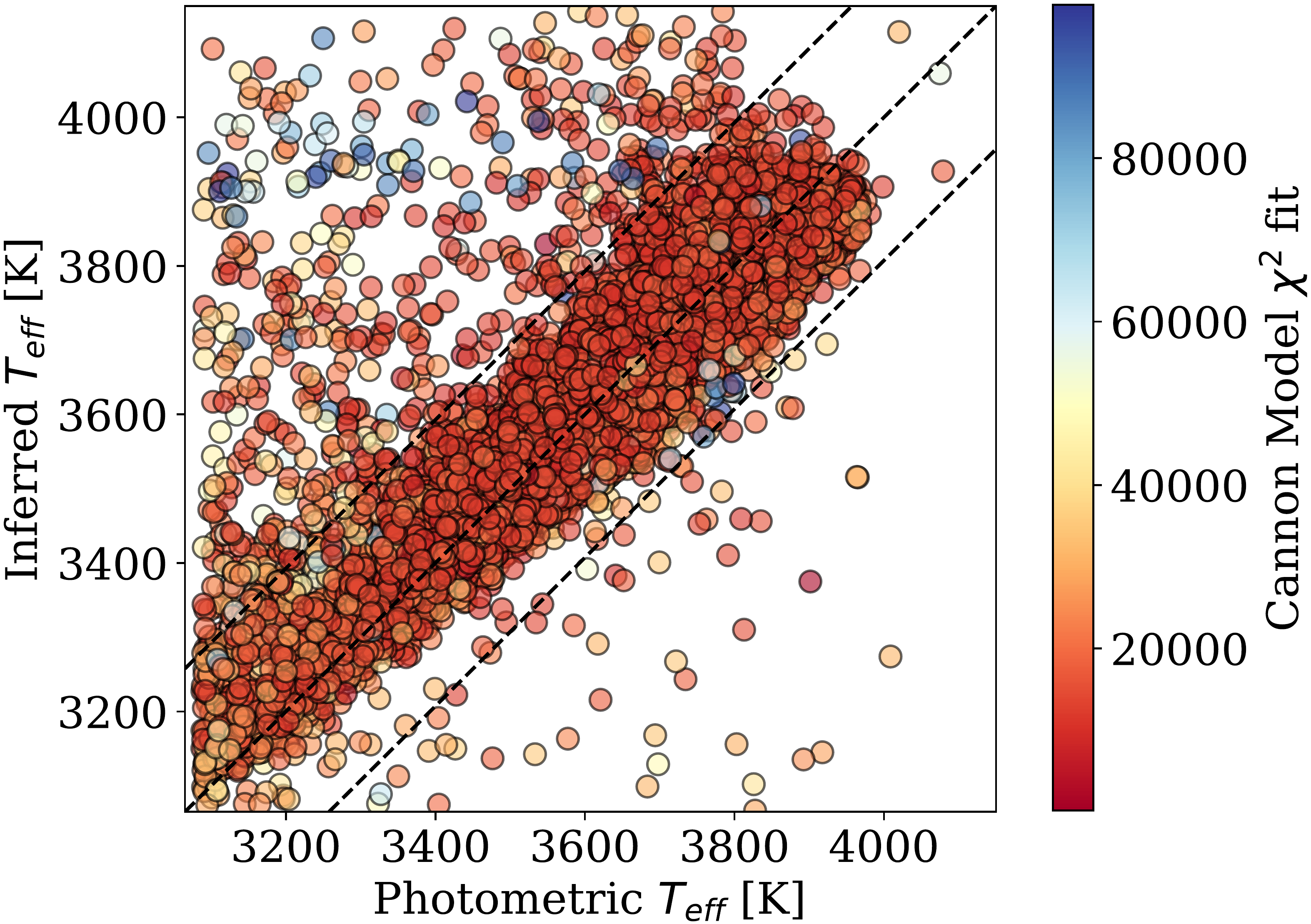} % second figure itself
\caption{$T_{\textrm{eff}}$ inferred from \emph{The Cannon} vs. $T_{\textrm{eff}}$ calculated from the \citet{curtis2020} relation for all $\sim$17,000 M dwarfs in our test set. The points are colored by the $\chi^{2}$ of the flux model spectral fits. The two dashed lines on either side of the 1-to-1 line mark the 2-$\sigma$ boundaries in $T_{\textrm{eff}}$ agreement.}
\label{fig:figure7}
\end{figure}

\begin{figure*}[t]
    \centering
    \includegraphics[width=0.988\textwidth]{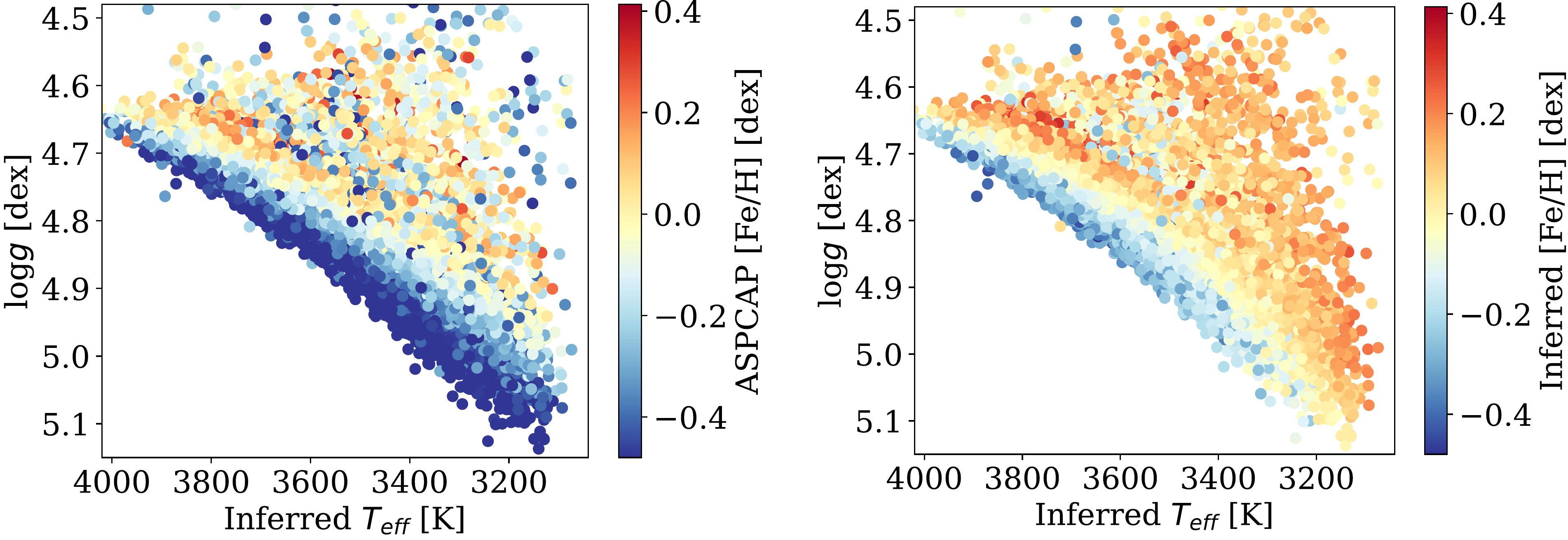} % second figure itself
\caption{Kiel diagrams of the M dwarfs in our test set (log $g$ from \citet{mann2019} versus the inferred $T_{\textrm{eff}}$ from our \emph{Cannon} model), colored by their [Fe/H] from the ASPCAP pipeline (not those of any potential FGK companions) (left), and inferred [Fe/H] from our \emph{Cannon} model (right).}
\label{fig:figure8}
\end{figure*}

\section{SDSS-V/MWM Test Sample}\label{sec:test_set}
We then use our FGK-M training set to infer detailed abundances for the largest sample of M dwarfs we can get from SDSS-V/MWM. Beginning with the entire SDSS-V/MWM catalog of $\sim$1 million stars, we select test set M dwarfs according to the type-color sequence relations from \citet{pecaut2013ApJS}. We then remove sources with negative \emph{Gaia} DR3 parallaxes, leaving us with $\sim$48,000 stars. Next, we implement additional cuts to select M dwarfs with high quality spectra, minimal binary contamination, and properties contained within the training set parameter space. We outline these cuts below:

\begin{enumerate}
    \item We apply a cut of SNR $\geq$ 50 to ensure that the test set spectra are sufficiently high quality for inferring high fidelity labels with \emph{The Cannon}.

    \item To remove contamination from binaries, we use \emph{Gaia} DR3 quantities to only retain M dwarfs that are well-fit by single star astrometric solutions. Specifically, we made cuts on the \emph{Gaia} Renormalised Unit Weight Error of RUWE $<$ 1.4, and set \texttt{non\_single\_star} = 0.

    \item We remove M dwarfs outside of the training set $T_{\textrm{eff}}$ boundaries with 3088 K $<$ $T_{\textrm{eff}}$ $<$ 4085 K. As mentioned in Section \ref{sec:sdssv_loocv}, we calculate $T_{\textrm{eff}}$ values from an empirical color-temperature relation \citep{curtis2020}, and use these photometric $T_{\textrm{eff}}$ values for this cut. %We implement a generous [Fe/H] cut of $-$0.7 $<$ [Fe/H] $<$ 0.7 dex using the ASPCAP [Fe/H] values. This exceeds the training set space ($-$0.56 $<$ [Fe/H] $<$ 0.31 dex), but we err on the side of being generous because ASPCAP is not accurate for M dwarfs.

    \item We calculate log $g$ values using an empirical relation based on \emph{Gaia} DR3 astrometry and $K$-band magnitudes \citep{mann2019}, and remove sources with log $g$ outside of 4$-$5.5 dex, the typical range that M dwarfs span (e.g., \citealt{casagrande2008}). Approximately 2\% of our sample lack $K$-band magnitudes, so we do not cut on log $g$ for these sources.

    \item After running our \emph{Cannon} model on the remaining M dwarfs in our test set following these cuts, we remove those with flux model spectral fit $\chi^{2}$ values $>$100,000 \citep{birky2020}.

    \item We also remove stars that lack sensible Hessian matrices from fits with our \emph{Cannon} model. This indicates that the log-likelihood space is flat around the critical values set by the best-inferred labels according our model, and as a consequence the model cannot identify the best direction to move in for optimizing the log-likelihood. In other words, the model fails to achieve a sensible fit.
    
\end{enumerate}

Following this set of cuts, we are left with $\sim$17,000 M dwarfs. Their flux model spectral fit $\chi^{2}$ values peak at $\sim$12,500 (reduced $\chi^{2}$ of $\sim$1.7), indicating that they are well-fit by the flux models. We show example flux model fits to four randomly selected M dwarfs from our test set that span a wide range of temperatures and metallicities in Figure \ref{fig:figureA4}.

We also use our flux model fits to explicitly show that our inferred abundances contain unique chemical information for each element. To do this, we selected four M dwarfs (SDSS ID = 79024069 overlaps with the set in Figure \ref{fig:figureA4}), of which two have very different inferred [Fe/H] vs. [Si/H], and the other two have very different inferred [O/H] vs. [Si/H]. We compare our flux models of these four M dwarfs with alternative flux models generated from substituting [Si/H] with [Fe/H] or [O/H], depending on which is more discrepant. It can be seen that our chosen flux models (red) outperform the alternative flux models (blue) in all cases, as shown by their fits to prominent Si and OH spectral features (Figure \ref{fig:figureA5}).

As another sanity check, we also compare the $T_{\textrm{eff}}$ values we infer from our model with $T_{\textrm{eff}}$ calculated from the empirical color-temperature relation derived in \citet{curtis2020}. The photometric $T_{\textrm{eff}}$ calculated from the empirical relation are not affected by reddening. More specifically, we find that $\sim$90\% of the M dwarfs have color extinction values E(BP$-$RP) $<$ 0.06, which corresponds to a maximum possible temperature shift of $\sim$60 K. Thus, our test set M dwarf \citet{curtis2020} relation $T_{\textrm{eff}}$ values are not strongly affected by reddening, and can be used for this comparison test. These $T_{\textrm{eff}}$ values and our inferred $T_{\textrm{eff}}$ values exhibit a clear 1-to-1 relation, with 89\% agreeing to within 100 K. In terms of significance, $T_{\textrm{eff}}$ agreement to within 2-$\sigma$ is at $\sim$193 K, and includes the bulk of sources near the 1-to-1 line (Figure \ref{fig:figure7}). The reason that some test set M dwarfs have large discrepancies between their $T_{\textrm{eff}}$ values from the \citet{curtis2020} relation versus from our model is likely because they fall outside the $T_{\textrm{eff}}$ and abundance parameter space spanned by our training set, and are being labeled with inaccurate $T_{\textrm{eff}}$ and abundance values that allow our model to achieve a good flux fit. We advise users of our catalog to treat such M dwarfs with caution, and include a flag (\textsf{temp\_agree}) on those with \citet{curtis2020} versus model $T_{\textrm{eff}}$ values discrepant beyond 2-$\sigma$. 

\begin{deluxetable*}{lllrrrrc}
\setlength{\tabcolsep}{1em}
\tablewidth{0.6\textwidth}
\tabletypesize{\footnotesize}
\tablewidth{0pt}
\tablecaption{M Dwarf Test Set Properties}
\tablecolumns{8}
\tablehead{
\colhead{\emph{Gaia} DR3 ID} &
\colhead{SDSS ID} &
\colhead{$T_{\textrm{eff}}$} &
\colhead{[Fe/H]} &
\colhead{[Mg/H]} &
\colhead{[Al/H]} &
\colhead{$\chi^{2}$} &
\colhead{\textsc{teff\_agree} \hspace{5mm}...} \\[-0.2cm]
\colhead{} &
\colhead{} &
\colhead{K} &
\colhead{dex} &
\colhead{dex} &
\colhead{dex} &
\colhead{} &
\colhead{\hspace{-8mm}} 
}
\startdata
421503353090268416 & 54381371 & 3815 $\pm$ 13.2 & 0.00 $\pm$ 0.02 & 0.03 $\pm$ 0.02 & 0.05 $\pm$ 0.02 & 8754 & True \\
421535342004177920 & 54381716 & 3830 $\pm$ 13.2 & $-$0.06 $\pm$ 0.02 & $-$0.05 $\pm$ 0.02 & $-$0.10 $\pm$ 0.02 & 12584 & True \\
421917594095590272 & 54386478 & 3382 $\pm$ 13.2 & 0.04 $\pm$ 0.02 & 0.14 $\pm$ 0.02 & 0.17 $\pm$ 0.02 & 16320 & True \\
422026445746556928 & 54387723 & 3603 $\pm$ 13.2 & 0.16 $\pm$ 0.02 & 0.16 $\pm$ 0.02 & 0.21 $\pm$ 0.02 & 19318 & True \\
422148079220822912 & 54389138 & 3792 $\pm$ 13.2 & $-$0.06 $\pm$ 0.02 & $-$0.04 $\pm$ 0.02 & $-$0.05 $\pm$ 0.02 & 59377 & True \\
422204253090333824 & 54389783 & 3280 $\pm$ 13.2 & $-$0.03 $\pm$ 0.02 & 0.03 $\pm$ 0.02 & 0.03 $\pm$ 0.02 & 23258 & True \\
422228171769143552 & 54390069 & 3578 $\pm$ 13.2 & 0.04 $\pm$ 0.02 & 0.10 $\pm$ 0.03 & 0.11 $\pm$ 0.03 & 8691 & True \\
422491023770086528 & 54392812 & 3755 $\pm$ 13.2 & $-$0.05 $\pm$ 0.02 & $-$0.01 $\pm$ 0.02 & $-$0.03 $\pm$ 0.02 & 59154 & True \\
422673886296537600 & 54394735 & 3674 $\pm$ 13.2 & $-$0.19 $\pm$ 0.02 & $-$0.10 $\pm$ 0.02 & $-$0.08 $\pm$ 0.02 & 11333 & True \\
422794420257500160 & 54396157 & 3505 $\pm$ 13.2 & 0.08 $\pm$ 0.02 & 0.13 $\pm$ 0.02 & 0.17 $\pm$ 0.02 & 20607 & True \\
422813352474333952 & 54396372 & 3815 $\pm$ 13.2 & $-$0.01 $\pm$ 0.02 & 0.03 $\pm$ 0.02 & 0.02 $\pm$ 0.02 & 21023 & True \\
423209863856063232 & 54400818 & 3861 $\pm$ 13.2 & 0.13 $\pm$ 0.02 & 0.13 $\pm$ 0.02 & 0.08 $\pm$ 0.02 & 16279 & True \\
423840051515323904 & 54408223 & 3816 $\pm$ 13.2 & 0.22 $\pm$ 0.02 & 0.20 $\pm$ 0.02 & 0.17 $\pm$ 0.02 & 26926 & True \\
424859165658264192 & 54421755 & 3755 $\pm$ 13.2 & 0.03 $\pm$ 0.02 & 0.12 $\pm$ 0.02 & 0.14 $\pm$ 0.02 & 11221 & True \\
424986674646826624 & 54423703 & 3345 $\pm$ 13.2 & 0.09 $\pm$ 0.02 & 0.12 $\pm$ 0.02 & 0.15 $\pm$ 0.02 & 10208 & True \\
425420638133142400 & 54429533 & 4119 $\pm$ 13.2 & 0.00 $\pm$ 0.02 & 0.00 $\pm$ 0.02 & $-$0.24 $\pm$ 0.02 & 16614 & False \\
425851195719880704 & 54434698 & 3769 $\pm$ 13.2 & 0.05 $\pm$ 0.02 & 0.08 $\pm$ 0.02 & 0.11 $\pm$ 0.02 & 15270 & True \\
426520355931927936 & 54444470 & 3716 $\pm$ 13.2 & $-$0.07 $\pm$ 0.02 & $-$0.04 $\pm$ 0.02 & $-$0.02 $\pm$ 0.02 & 8546 & True \\
426617864561108736 & 54446189 & 3803 $\pm$ 13.2 & 0.05 $\pm$ 0.02 & 0.03 $\pm$ 0.02 & $-$0.00 $\pm$ 0.02 & 25069 & True \\
426622198194555648 & 54446259 & 3855 $\pm$ 13.2 & 0.07 $\pm$ 0.02 & 0.08 $\pm$ 0.02 & 0.05 $\pm$ 0.02 & 10447 & True \\
427277335326484096 & 54456095 & 3380 $\pm$ 13.2 & 0.17 $\pm$ 0.02 & 0.22 $\pm$ 0.02 & 0.29 $\pm$ 0.02 & 15818 & True \\
427880245651962880 & 54464028 & 3904 $\pm$ 13.2 & 0.02 $\pm$ 0.02 & 0.05 $\pm$ 0.02 & 0.02 $\pm$ 0.02 & 17876 & True \\
428015073265343872 & 54465640 & 3610 $\pm$ 13.2 & $-$0.17 $\pm$ 0.02 & $-$0.08 $\pm$ 0.02 & $-$0.08 $\pm$ 0.02 & 17593 & True \\
428216833650287232 & 54467942 & 3509 $\pm$ 13.2 & $-$0.01 $\pm$ 0.02 & $-$0.00 $\pm$ 0.03 & 0.00 $\pm$ 0.03 & 11967 & False \\
428262703901618432 & 54468635 & 3713 $\pm$ 13.2 & 0.03 $\pm$ 0.02 & 0.05 $\pm$ 0.02 & 0.03 $\pm$ 0.02 & 19496 & True \\
428342933890361216 & 54469672 & 3944 $\pm$ 13.2 & $-$0.09 $\pm$ 0.02 & $-$0.01 $\pm$ 0.02 & $-$0.06 $\pm$ 0.02 & 10746 & True \\
428347778613655424 & 54469716 & 3544 $\pm$ 13.2 & 0.05 $\pm$ 0.02 & 0.06 $\pm$ 0.02 & 0.09 $\pm$ 0.02 & 37802 & True \\
428489409455550336 & 54471368 & 3614 $\pm$ 13.2 & 0.02 $\pm$ 0.02 & 0.02 $\pm$ 0.02 & 0.04 $\pm$ 0.02 & 8269 & True \\
428538818759899648 & 54472086 & 3841 $\pm$ 13.2 & 0.13 $\pm$ 0.02 & 0.13 $\pm$ 0.02 & 0.10 $\pm$ 0.02 & 10299 & True \\
428548645644559104 & 54472241 & 3187 $\pm$ 13.2 & 0.10 $\pm$ 0.02 & 0.14 $\pm$ 0.02 & 0.17 $\pm$ 0.02 & 33632 & True \\
& & & ...\\
\enddata
\tablecomments{This table lists the properties of the 16,590 M dwarfs in our SDSS-V/MWM test set. The properties consist of all labels inferred from our implementation of \emph{The Cannon} ($T_{\textrm{eff}}$, abundances for all elements of interest, and flux model spectral fit $\chi^{2}$ values). The errors on the $T_{\textrm{eff}}$ and abundance labels are the scatter in labels from resampling from flux errors 10 times for each M dwarf. We also include a \textsc{teff\_agree}) flag describing whether the photometric and \emph{Cannon}-inferred $T_{\textrm{eff}}$ agree to within 2-$\sigma$. We only list a subset of the abundances in this table, but the full set is provided in the downloadable version.
\vspace{1.5mm}
\newline(This table is available in its entirety in machine-readable form.)}
\end{deluxetable*} \label{tab:table2}

Using our $T_{\textrm{eff}}$ inferred from \emph{The Cannon} and log $g$ calculated from the \citet{mann2019} relation, we plot a Kiel diagram for our test set M dwarfs. The M dwarfs are colored by our inferred [Fe/H], which exhibit the evolutionary tracks we expect as dictated by varying stellar metallicities (e.g., \citealt{hejazi2022}) (Figure \ref{fig:figure8}, right panel). Our inferred [Fe/H] values also appear more reasonable than the ASPCAP [Fe/H] values, which are unrealistically metal-poor ($>$80\% have ASPCAP [Fe/H] $<$ 0 dex) (Figure \ref{fig:figure8}, left panel). This indicates that our M dwarf abundances inferred from \emph{The Cannon} are reliable, and an improvement on those from the current SDSS-V ASPCAP pipeline. 

In order to quantify the inferred $T_{\textrm{eff}}$ and abundance label uncertainties of the $\sim$17,000 M dwarfs in our test set, we estimate the scatter between labels from repeat APOGEE observations of the same stars. We identify $\sim$500 stars in our sample with two visit spectra from APOGEE, denoted $A$ and $B$, and define a quantity $Z_{A,B}$ for each star and label:

\begin{equation} \label{eq:equation11}
Z_{A,B} = \frac{\ell_{A} - \ell_{B}}{\sqrt{\sigma_{A}^{2} + \sigma_{B}^{2} + 2(\sigma_{inflate}^{2})}}, 
\end{equation}

\noindent where $\ell_{A}$ and $\ell_{B}$ are the inferred values for a particular label ($T_{\textrm{eff}}$ or an elemental abundance) using spectrum $A$ or $B$, and $\sigma_{A}$ and $\sigma_{B}$ are the associated scatter derived from our \emph{Cannon} model covariance matrices. We fit for $\sigma_{inflate}$, which can be thought of as the additional scatter we need to account for the difference in inferred labels from different APOGEE visit spectra $A$ and $B$. The $\sigma_{inflate}$ term is multiplied by two to account for the two visit spectra. 

We fit for $\sigma_{inflate}$ factors so that the 13 distributions of $Z_{A,B}$ (corresponding to our 13 labels) from our sample of $\sim$500 stars are normal distributions centered at 0, with standard deviations of 1. The resultant $\sigma_{inflate}$ values range from 0.016$-$0.025 dex. To derive the final $T_{\textrm{eff}}$ and abundance label uncertainties for each star, we add these $\sigma_{inflate}$ factors in quadrature with each star's scatter for that label from their individual covariance matrix. The median values of the resultant $T_{\textrm{eff}}$ and abundance errors across all elements is 13 K and 0.018$-$0.029 dex, respectively. We provide these uncertainties and all other properties of our final catalog containing 16,590 M dwarfs in Table \ref{tab:table2}.

\section{Conclusions}\label{sec:conclusion}
Using \emph{The Cannon}, we constructed a data-driven model for inferring M dwarf abundances across a wide set of elements. We apply our model to 16,590 M dwarfs in SDSS-V/MWM, and provide a catalog of their inferred $T_{\textrm{eff}}$ and abundances. We anticipate that this catalog will be invaluable for star and planet formation investigations with SDSS-V data.

We note that our M dwarf model parameter space spans $-$0.56 $<$ [Fe/H] $<$ 0.31 dex, and is not suitable for inferring abundances for M dwarfs outside this metallicity range. We do not expect that many M dwarfs in SDSS-V/MWM will be more metal-poor than $-$0.56 dex, or more metal-rich than 0.31 dex given that M dwarfs are often quite old. Still, users should be confident that their M dwarfs of interest fall within this metallicity range before using our catalog. In the future it may be possible to expand our model parameter space with a larger, more diverse FGK-M training set, but this will require identification of more FGK-M binaries in SDSS-V/MWM. For now, the metallicity range spanned by our catalog is about as wide as any other existing method for inferring M dwarf abundances in large stellar samples (e.g., \citealt{birky2020}).
%For example, the \citet{birky2020} catalog of M dwarf $T_{\textrm{eff}}$ and [Fe/H] spans $-$0.5 $<$ [Fe/H] $<$ 0.5 dex, and includes only two $>$0.3 dex sources in their training set.

Our validation tests with the \citet{wanderley2023} Hyades M dwarf abundances and the \citet{souto2022} M dwarf sample indicate that our M dwarf model is robust. The inferred metallicities for our final sample of 16,590 M dwarfs also exhibit the evolutionary tracks we expect according to the stellar metallicities. As more M dwarfs are observed through SDSS-V/MWM, our model can be used to infer their detailed abundances as their spectra become available. Additionally, our model may be compatible with M dwarf spectra collected with other instruments/surveys, but this may require additional processing, e.g., spectra from an instrument with higher resolution compared to APOGEE must be artificially degraded to match the SDSS-V/MWM training set. The spectra must also have a wavelength range encompassed by APOGEE, i.e., $H$-band. Expanding our model to work with M dwarf spectra from other instruments could be quite fruitful, but we leave these investigations for future studies.

\section*{Acknowledgements}
%\begin{acknowledgments} % uncomment for line numbers
We thank Adrian Price-Whelan, David Hogg, F\'{a}bio Wanderley, Ilija Medan, Zach Way, Shubham Kanodia, and Soichiro Hattori for productive conversations. D.S. thanks the National Council for Scientific and Technological Development—CNPq process No. 404056/2021-0. We also thank the anonymous referee for a helpful report. 

Funding for the Sloan Digital Sky Survey V has been provided by the Alfred P. Sloan Foundation, the Heising-Simons Foundation, the National Science Foundation, and the Participating Institutions. SDSS acknowledges support and resources from the Center for High-Performance Computing at the University of Utah. SDSS telescopes are located at Apache Point Observatory, funded by the Astrophysical Research Consortium and operated by New Mexico State University, and at Las Campanas Observatory, operated by the Carnegie Institution for Science. The SDSS web site is \url{www.sdss.org}.

SDSS is managed by the Astrophysical Research Consortium for the Participating Institutions of the SDSS Collaboration, including the Carnegie Institution for Science, Chilean National Time Allocation Committee (CNTAC) ratified researchers, Caltech, the Gotham Participation Group, Harvard University, Heidelberg University, The Flatiron Institute, The Johns Hopkins University, L'Ecole polytechnique f\'{e}d\'{e}rale de Lausanne (EPFL), Leibniz-Institut f\"{u}r Astrophysik Potsdam (AIP), Max-Planck-Institut f\"{u}r Astronomie (MPIA Heidelberg), Max-Planck-Institut f\"{u}r Extraterrestrische Physik (MPE), Nanjing University, National Astronomical Observatories of China (NAOC), New Mexico State University, The Ohio State University, Pennsylvania State University, Smithsonian Astrophysical Observatory, Space Telescope Science Institute (STScI), the Stellar Astrophysics Participation Group, Universidad Nacional Aut\'{o}noma de M\'{e}xico, University of Arizona, University of Colorado Boulder, University of Illinois at Urbana-Champaign, University of Toronto, University of Utah, University of Virginia, Yale University, and Yunnan University.

This work made use of data from the European Space Agency (ESA) mission Gaia (\url{https://www.cosmos.esa.int/gaia}), processed by the Gaia Data Processing and Analysis Consortium (DPAC, \url{https://www.cosmos.esa.int/web/gaia/dpac/consortium}). Funding for the DPAC has been provided by national institutions, in particular the institutions participating in the Gaia Multilateral Agreement.
\software{\texttt{numpy} \citep{numpy}, \texttt{matplotlib} \citep{matplotlib}, \texttt{pandas} \citep{pandas}, \texttt{scipy} \citep{scipy}, \texttt{scikit-learn} \citep{scikit-learn}, \texttt{astropy} \citep{astropy:2013, astropy:2018}, \texttt{thecannon} \citep{ness2015,casey2016}}
%\end{acknowledgments}

\bibliography{mybib}{}
\bibliographystyle{aasjournal}

\appendix
\setcounter{figure}{0}                       
\renewcommand\thefigure{A.\arabic{figure}}

\begin{figure*}[t]
    \centering
    \includegraphics[width=0.988\textwidth]{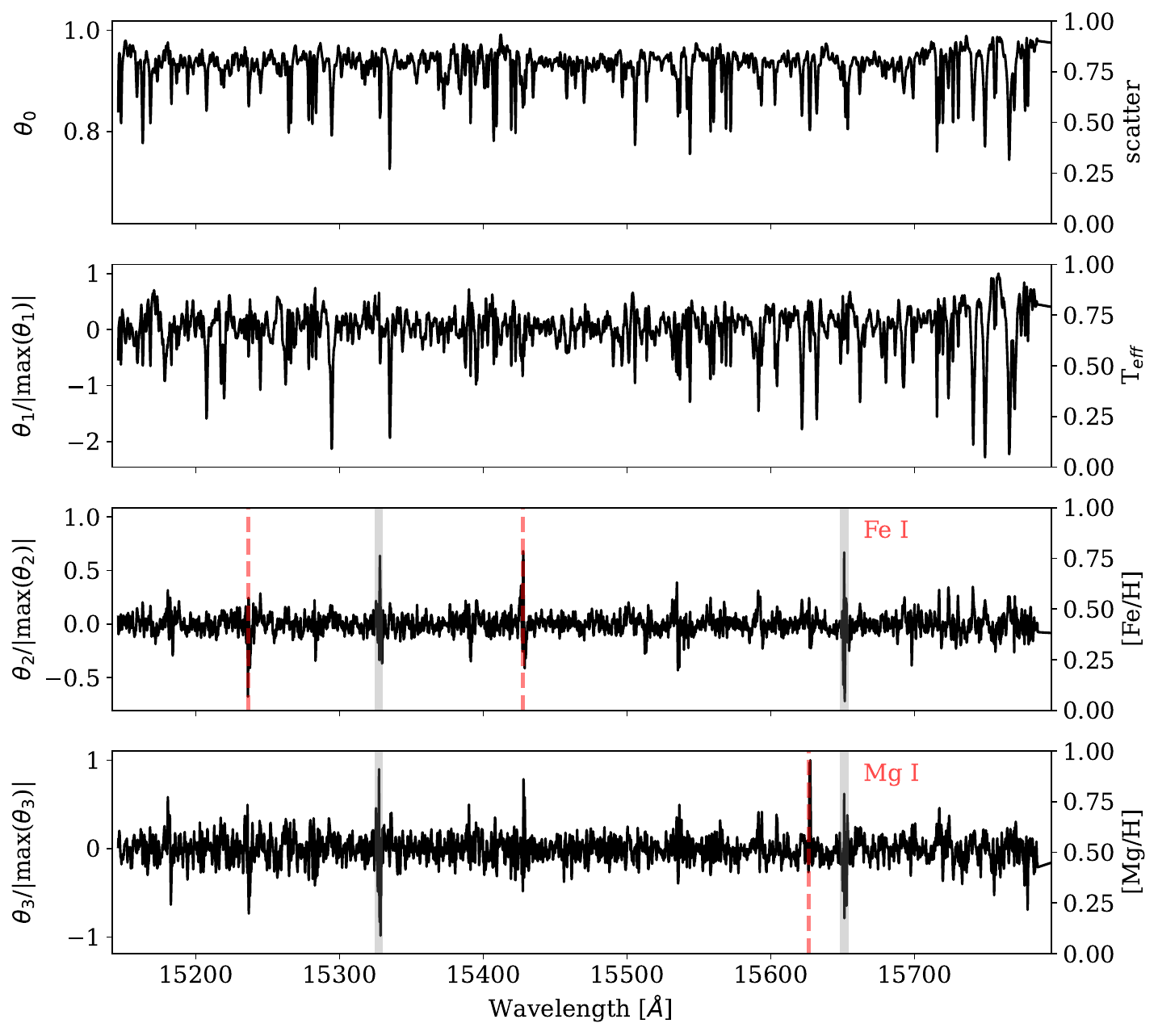}
\caption{The scatter and normalized flux coefficients for the first four labels of our \emph{Cannon} model trained on FGK-M binaries. The wavelength pixels with large coefficient amplitudes correspond to locations in the spectra that contain the most information for that particular label. We highlight a few prominent absorption features whose locations match high amplitude pixels. We also highlight two wavelength pixels with poor sky subtraction in gray, which should be disregarded (these wavelength pixels are assigned high intrinsic model scatter $s^{2}_{j}$, so they do not affect the results). For ease of viewing, we only show the wavelength range covered by the first APOGEE chip.}
\label{fig:figureA1}
\end{figure*}

\begin{figure*}[t]
    \centering
    \includegraphics[width=0.988\textwidth]{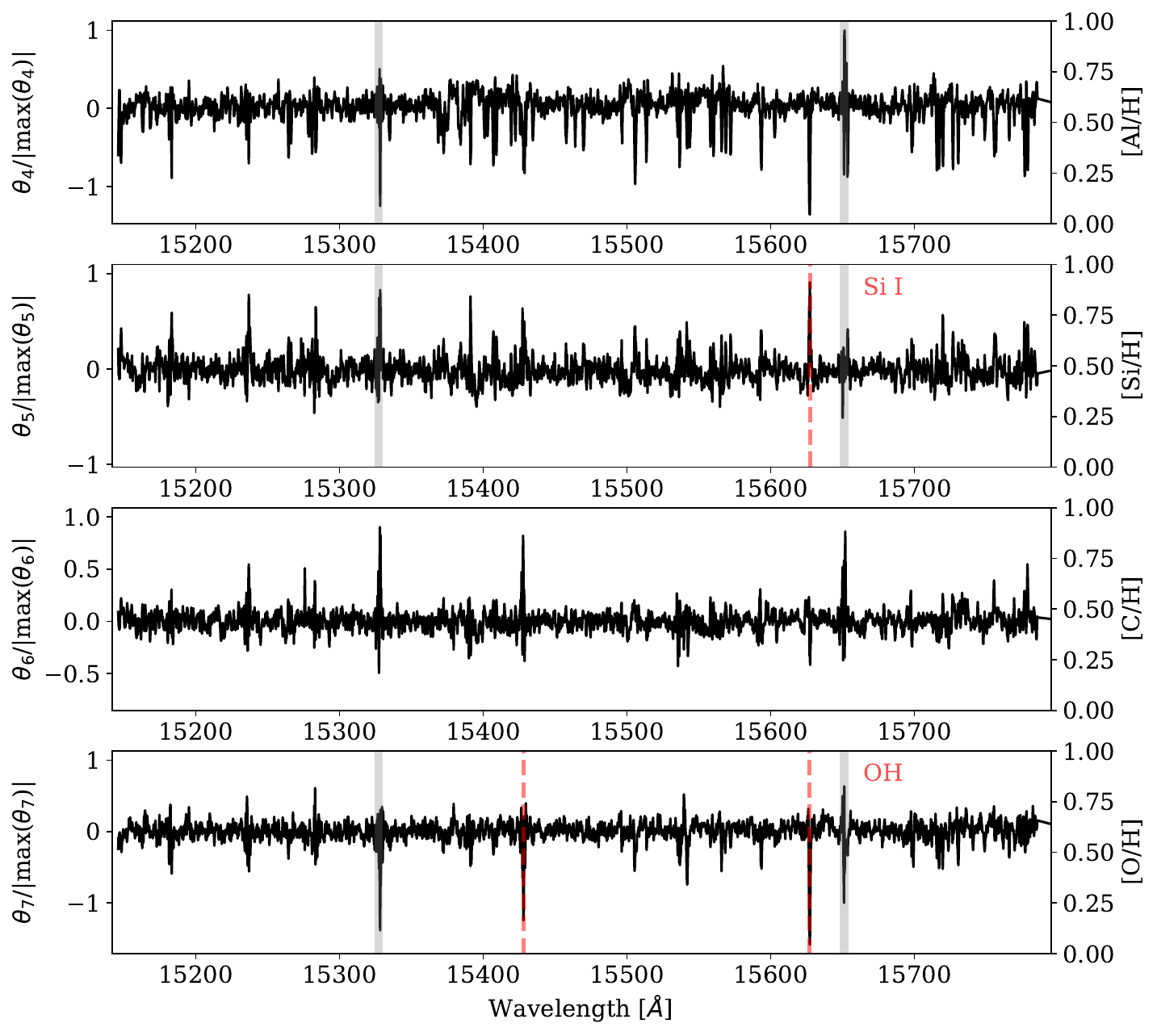}
\caption{The scatter and normalized flux coefficients for the next four labels of our \emph{Cannon} model.}
\label{fig:figureA2}
\end{figure*}

\begin{figure*}[t]
    \centering
    \includegraphics[width=0.988\textwidth]{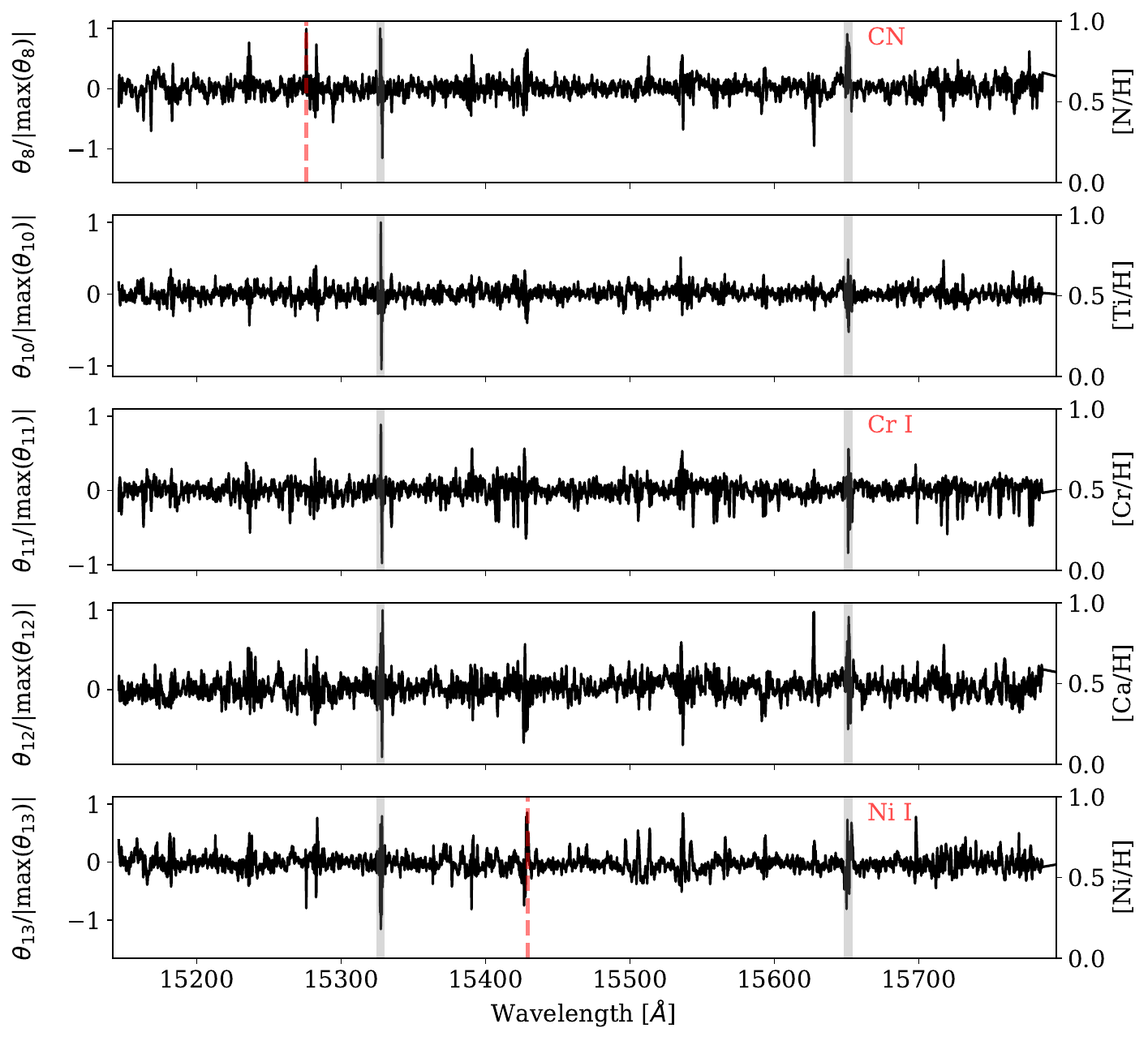}
\caption{The scatter and normalized flux coefficients for the final five labels of our \emph{Cannon} model.}
\label{fig:figureA3}
\end{figure*}

\begin{figure*}[t]
    \centering
    \includegraphics[width=0.9\textwidth]{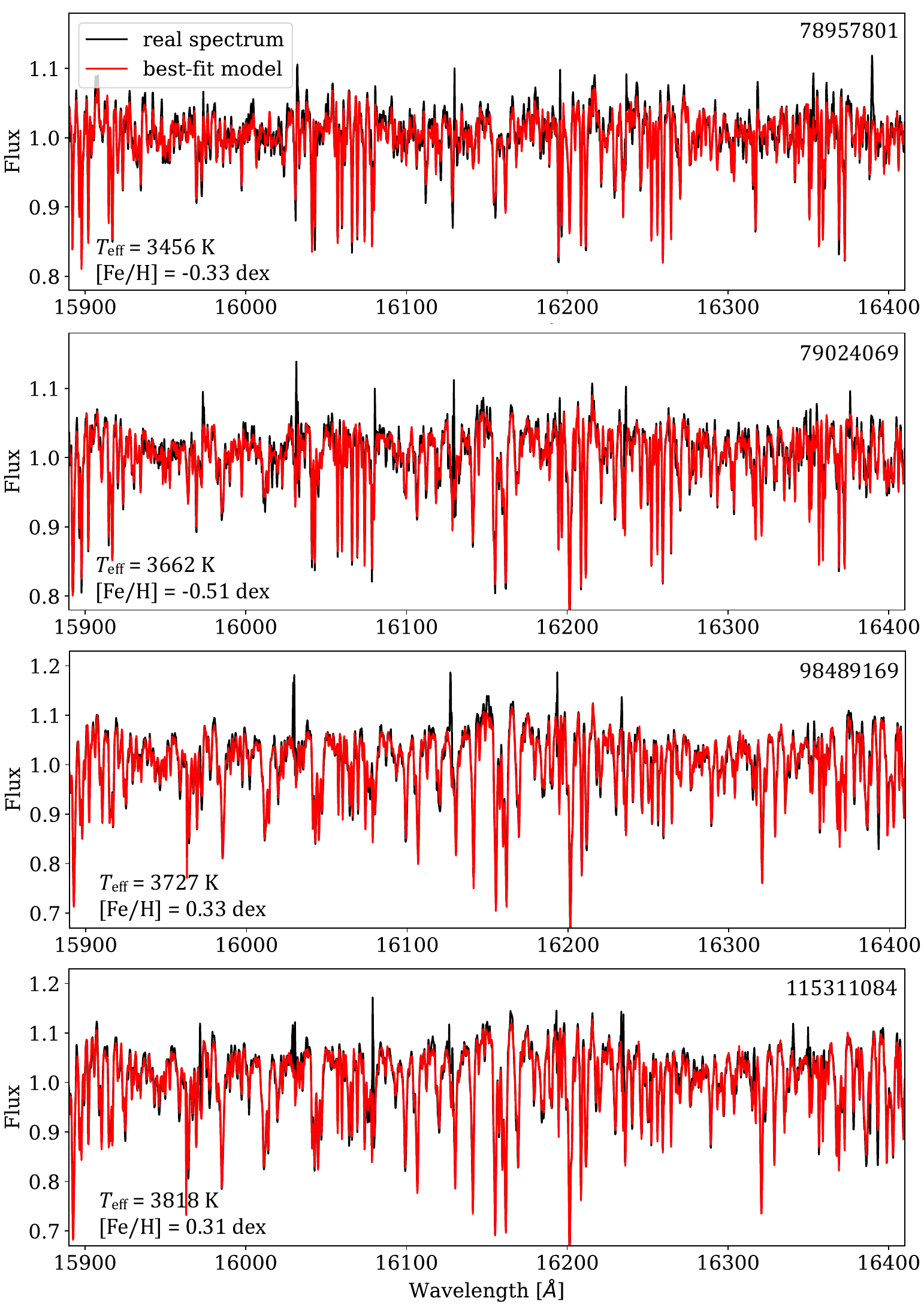}
\caption{Plots of flux model fits from our \emph{Cannon} model to four randomly selected M dwarfs from our test set that span a wide range of temperatures and metallicities. We show the wavelength range of the middle APOGEE chip. The model fits are in red, and the real spectra are in black. The SDSS IDs of each M dwarf are provided in the top right corner of each panel.}
\label{fig:figureA4}
\end{figure*}

\begin{figure*}[t]
    \centering
    \includegraphics[width=0.988\textwidth]{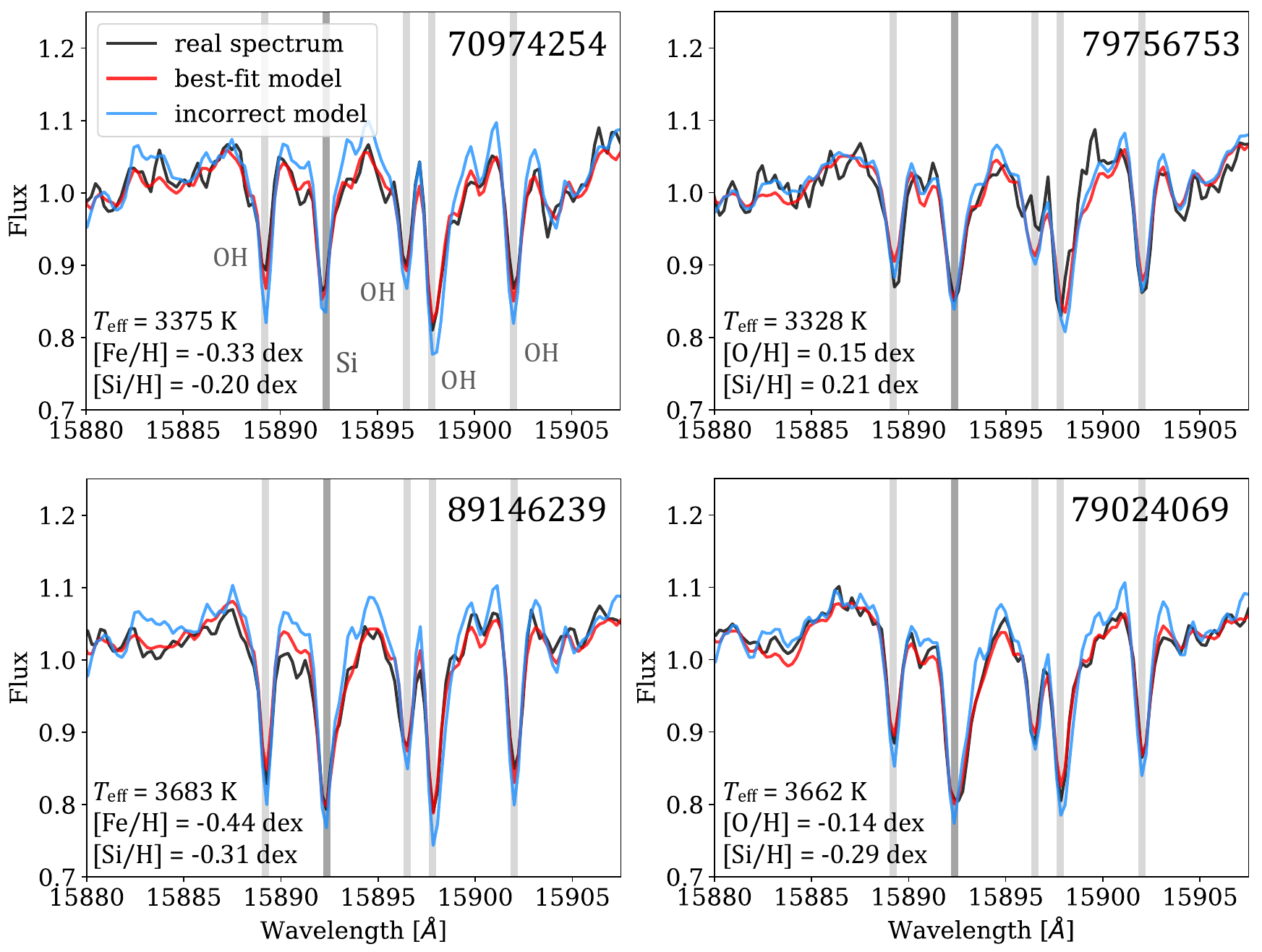}
\caption{Plots of our flux model fits to four M dwarfs in our test set (SDSS IDs provided in the top right corner of each panel), of which two have very different inferred [Si/H] and [Fe/H] values (left column), and the remaining two have very different interred [Si/H] and [O/H] values (right column). The spectra (black) are compared to our chosen flux models (red), and alternative flux models where we substitute [Si/H] with either [Fe/H] or [O/H] depending on which is most discrepant. Our chosen models outperform these alternative models for all M dwarfs, especially in the region we highlight which contains prominent Si (dark gray line), and OH (lighter gray lines) absorption features.}
\label{fig:figureA5}
\end{figure*}

\end{document}